 \documentclass[aps,prd,superscriptaddress,floatfix]{revtex4-1}
\usepackage{color}
\usepackage{amssymb}
\usepackage[normalem]{ulem}
\usepackage{graphicx,color}
\usepackage{amsmath}
\usepackage{subfigure}
\usepackage[normalem]{ulem}
\usepackage{booktabs}
\usepackage{xcolor}
\usepackage{epsfig,graphics,color,graphicx,amsmath}

\colorlet{darkgreen}{green!50!black}
\colorlet{brightyellow}{yellow!75!red}
\colorlet{orange}{red!50!yellow}
\colorlet{darkblue}{blue!60!black}
\colorlet{darkred}{red!80!black}

\usepackage{soul}

\usepackage{mathtools}
\usepackage{mathrsfs}
\usepackage{bm}
\usepackage{relsize}
\usepackage{indentfirst}

\usepackage{xcolor}

\usepackage{amsmath}   
\usepackage{hyperref}  

\usepackage{cleveref}  

\usepackage{verbatim} 
\usepackage{amssymb,amsmath,multirow,epsfig,graphicx,color}

\usepackage{epsfig}
\usepackage{graphicx,amsmath}
\def\be{\begin{eqnarray} &&}

\def\ee{\end{eqnarray}}

\newcommand\ba{\begin{eqnarray}}
\newcommand\ea{\end{eqnarray}}

\newcommand{\bas}{\begin{eqnarray*}}
	\newcommand{\eas}{\end{eqnarray*}}

\newcommand{\bno}{\begin{eqnarray*}}
	\newcommand{\eno}{\end{eqnarray*}}

\def\sl

\usepackage{hyperref}
\usepackage{url}
\usepackage{hyperref}
\hypersetup{
	colorlinks=true,  
	linkcolor=blue,   
	filecolor=magenta,      
	urlcolor=blue,    
	citecolor=blue,   
} 
\usepackage{braket}
\usepackage{slashed}

\begin{document}

	\hfill \hspace{-1cm} LFTC - 26.08/117
	

	\title{ Behavior of the zero modes in the Light-front for pseudo scalar particles: pion case }
	\author{J.~P.~B.~C.~de Melo}
	\email{jpachecodm@gmail.com \ and \
		 joao.mello@cruzeirodosul.edu.br} 
	\affiliation{Laborat\'orio de F\'\i sica Te\'orica e Computacional - LFTC, \\
		Universidade Cruzeiro do Sul and Universidade Cidade de S\~ao Paulo (UNICID),
		 01506-000 S\~ao Paulo, Brazil}
	\affiliation{Instituto Tecnol\'ogico de Aeron\'autica, 12.228-900,
		 S\~ao Jos\'e dos Campos, SP, Brazil}	
	\date{\today}

		\begin{abstract}
	
		Light-front Quantum Field Theory (LFQFT) stands as an 
		optimal framework for 
		\linebreak 
		delving into the intricate substructure of hadronic
		bound states. In this study, we employ a 
		\linebreak  
		Light-Front Constituent Quark Model (LFCQM) 
		to investigate the pion's internal structure. This model has previously
		demonstrated significant success in describing the properties of pseudoscalar mesons.
		A crucial consideration within the light-front formalism is the dependence on the
		chosen
		\linebreak 
		reference frame, stemming from the explicit breaking of covariance. Herein, we meticulously
		analyze the 
		contribution of zero modes to the pion's electromagnetic current. 
		Subsequently, we explore the impact of these zero modes on observable quantities,
		including electromagnetic form factors, the pion's electromagnetic radius,
		and its weak decay constant, $f_\pi$.
		
		Furthermore, we comprehensively investigate the behavior of zero modes and their
		dependence on both the momentum transfer, $Q^2$, and the model's intrinsic parameters, 
		such as constituent quark masses ($m_{(q,\bar{q})}$) and the regulator mass ($m_R$).
		
		 Our findings reveal 
		a substantial dependence of light-front zero modes on the specific frame employed 
		for the calculation of the electromagnetic current. Interestingly, we observe 
		that the maximal contribution of the zero mode occurs at a specific 
		four-momentum transfer, denoted as $Q^2 = -q^2 = q_Z^{2(Max.)}$.	
	\end{abstract}
	
	\maketitle

\section{Introduction} 
After the seminal paper by Paul Dirac \cite{Dirac1949}, it is possible to 
describe the quantum field theory, with approaches different from the equal-time 
relativistic quantum mechanics ~\cite{Zuber}. Between other, Paul Dirac proposed
the 
\linebreak 
light-front form for the relativistic quantum field theory, or, was usual 
denominate, light-front or light-cone~\cite{Brodsky1998,Brodsky2004}. 
Today, it is believed, that strong interactions are governed by the Quantum Chromodynamics~(QCD),
the quantum relativistic theory of the quarks and gluons, the basic degrees of freedom in
the nature~\cite{PDG2026}. In principle, it is possible, to described with that theory,
hadronic bound states, it is composite systems with quarks and
 gluons~\cite{Brodsky1998,Pagels1978,Pagels1979}. 
 In quantum field theory, as usual, the electromagnetic form factors provide informations about
the hadronic structure. These form factors are defined as scalar functions of
Lorentz structures, and comes from the decomposition" of the matrix elements
from the hadronic electromagnetic currents~\cite{Zuber}.  
Due to discrete symmetries and
the covariance of the electromagnetic current,
pseudoscalar particles are typically described by 
a single form factor when both the initial and final
mesons are on their mass shells~\cite{Yabusaki2015,Moita2021}, 
 as is the case in the present work.
However, recent studies have explored scenarios where the initial 
or final meson is off-shell. In such cases, off-shell effects emerge,
requiring the introduction of two form factors to maintain
the full relativistic description of the vertex~\cite{Choi2019,Choi2020,Leao2024}.

	The study of pseudoscalar mesons is an important source of information on bound states
		 using QCD based models, which can lead to many distinct
		  approaches \cite{Sterman1997,Burkert2022,Horn2016}.

	Among these approaches are lattice calculations
	 \cite{Dudek2015,Oliveira2019,Alexandrou2022,Delmar2024,Alexadrou2024},
	  which discretize space-time into a finite grid to non-perturbatively solve quantum chromodynamics (QCD).
On the other hand, there is a non-perturbative development based on QCD, 
in terms of the Dyson-Schwinger and Bethe-Salpeter equations, which constitutes a 
covariant formalism \cite{Roberts1994,Cloet2014,Xu2024,Hernandez2023,Ding2026} that 
encompasses both confinement and chiral symmetry breaking.
In complementing these continuum and lattice aproaches,
 the Nambu–Jona-Lasinio (NJL) model provides a non-perturbative effective framework 
 that captures key features of QCD, in special the dynamics  of the 
  chiral symmetry breaking,  and the Goldstone-boson nature of the pion
   \cite{Klevansky1992,Arriola2002,Hutauruk2016,Bentz2025}.  
   There are many phenomenological and theoretical studies of the pion, particularly regarding its static and dynamic properties~\cite{Frederico1994,Ebert2005,Krutov2009}, as well as experimental ones~\cite{Brauel1977,Amendolia1986,Baldini1998,Baldini2000,Volmer2000,Horn2006,
   	Tadevosyan2007,Horn2007,Blok2008,Huber2008}.

The light-front approach is the ideal formalism for studying bound hadronic systems, 
whether for spin-zero pseudoscalar mesons
 \cite{Terentev1976,Brodsky1980,Frederico1994,Moita2021,Krutov2009,JVary2018,Meijian2019,Lan2019a,Lan2019b,Liu2023},
 spin-1 mesons \cite{Jaus1999,Jaus2003,Choi2004,Choi1999,Liu2023,Xu2025,Sun2017,Sun2017b}, 
 as well baryons \cite{deAraujo1995,deMelo2008,deMelo2008rj,Xu2023,Kaur2024,Wu2026,Dahiya2026}.

In the present work, the pion structure is investigated using a Light-Front
 Constituent Quark Model (LFCQM) with both symmetric \cite{deMelo2002,Moita2021} 
 and non-symmetric vertex models \cite{deMelo1999,Jurandi2023a}. 
 The LFCQM approach enables the derivation of light-front hadronic wave functions and the study of various hadronic properties.

 However, the primary objective of this work is to identify patterns within the non-valence contributions to the electromagnetic current when the light-front formalism
  is applied; it is important to note that such contributions do not arise in equal-time quantum theory. These non-valence contributions significantly affect the calculation of hadronic observables, particularly electromagnetic form factors. Their exclusion in the light-front approach leads to a violation of essential
   symmetries, such as rotational symmetry
  in the electromagnetic current elements   
\cite{Choi2004,Meijian2019,Zhang2025,deMelo2026}.  
With  the light-front approach, 
the electromagnetic current for pseudoscalar particles exhibits a dependence on the adopted reference frame, which leads to a violation of rotational symmetry. Consequently, contributions to the electromagnetic form factor arise beyond the valence sector. This theme has been a frequent subject in recent literature, concerning both 
pseudoscalar particles and spin-1 vector particles.

This work is organized into three additional sections. Section \ref{models} presents the electromagnetic current for pseudoscalar particles, as well as the light-front models employed here to calculate the elastic 
electromagnetic form factor of the pion. Section \ref{results} details the numerical calculations and the results obtained in this study. In section \ref{summary}, we present our conclusions regarding the role of zero modes in the calculation of the electromagnetic current across different reference frames, and how this affects the extraction of observables for the pion.


\section{The electromagnetic current and the Light-front models}
\label{models}
The electromagnetic properties of bound states are a fundamental branch of the hadronic physics; in the present section, we implement a simplified framework to establish the spin coupling
 for the quark-antiquark system within the meson structure, derived from an effective Lagrangian.
  It is important to mention that meson vertex function will be subsequently incorporated,
   introducing a 
 mass scale determined by the weak decay constant.

The interaction between quark fields and the pseudoscalar mesonic field is formulated
 under the flavor symmetry group $SU(2)$, and is described by the following
 effective interaction Lagrangian density:
\begin{equation}
\mathcal{L}_I = -i g \bar{\Psi} M_{\mathrm{SU}(2)} \gamma^5 \Psi
 \equiv -i g \sum_{i=1}^{3} \left( \bar{\Psi} \tau_i \gamma^5 \Psi \right) \varphi^i
\label{eq:lagrangiana}
\end{equation}

In this formulation, the summation index runs from $1$ to $3$, corresponding to 
the three generators of the $SU(2)$ group. These generators are represented by the
 isospin operators $\tau_i = \frac{1}{2}\sigma_i$ (where $\sigma_i$ are 
 the standard Pauli matrices), ensuring consistency with the algebraic structure
  of the isospin group. The quark doublet is defined 
  by $\Psi = \begin{pmatrix} u \\ d \end{pmatrix}$, while $\varphi^i$
   represents the components of the mesonic field. The mesonic field 
   matrix, $M_{\mathrm{SU}(2)}$, is explicitly constructed through 
   the scalar product of the generators and the adjoint fields $\varphi^i$, 
   where the physical fields of the charged pions and the $\pi^0$ are defined by:

\begin{align}
M_{\mathrm{SU}(2)} = \sum_{i=1}^{3} \tau_i \varphi^i &= 
\frac{1}{2}\begin{pmatrix}
\pi^0 &  2 \pi^{+} \\ \nonumber 
2 \pi^{-} & -\pi^0
\end{pmatrix} \\[10pt]
\pi^{+} &=  \operatorname{Tr}\left[M_{\mathrm{SU}(2)}\left(\tau_1 +  i \tau_2\right)\right]
\end{align}

In the present work, we use the effective Lagrangian from the 
reference \cite{Moita2021}, but now, with two models for the vertex 
bound sate quark-antiquark \cite{deMelo1999,deMelo2002}, with the 
following  expression for the Beth-Salpeter amplitude. 
\begin{equation}
\Psi_M(k, p)=S_q(k) \gamma^5 g \Gamma_M(k, p) \lambda_M S_q(k-p)
\end{equation}
The quark propagador quark and anti-quark are given below:
\begin{equation}
S_q(k) = i \frac{\slashed{k} + m_q}{\left( k^2 - m_q^2 + i \epsilon \right)},
\ m_{q} \ \leftrightarrow \ m_{\bar q}
\end{equation}

The  matrix for the quarks masses is, $diag[m_q]=[m_u,m_d]$.
 For the two $\pi-q\bar{q}$ vertex functions utilized in the present work,
 was used to calculate the electromagnetic properties for the pion,
  and compared with the experimental data
  \cite{deMelo1999,deMelo2002,Yabusaki2015,Jurandi2023a,daSilva2012}.  
The results obtained with these vertex models are in good agreement
 with experimental data, both, for  the electromagnetic factor, as well
  as the charge radius and the electroweak constant of the pion.
The first vertex is given by a symmetric function 
by the change momentum , quark anti-quark,  
as used before in \cite{deMelo2002,Yabusaki2015,Moita2021}.
The second vertex utilized, is the non-symmetric vertex function, 
previously utilized \cite{deMelo1999,  daSilva2012}.
Both models vertex functions, as written  in the following expression below:

\begin{equation}
g\Gamma^{(i)}(k, p) =  \frac{N}{k^2 - m_R^2 + i\epsilon} + \eta 
 \frac{N}{(p-k)^2 - m_R^2 + i\epsilon}
~,
\label{unified}
\end{equation}
where $\eta = 1$ corresponds to the symmetric vertex model 
$\Gamma^{\tiny (SY)}$
~\cite{deMelo2002}
, and $\eta = 0$ corresponds to the non-symmetric vertex model $\Gamma^{\tiny (NSY)}$
~\cite{deMelo1999}, with the change $k \rightarrow (p-k)$.

In the vertex functions,Eq.(\ref{unified}); $m_{R}$ acts as a regulator mass; 
 which is not determined by the model itself, is introduced to ensure  
the amplitudes convergence and to signify soft effects at short ranges. 
This value is calibrated using the experimental value for the electroweak
decay constant for the pion (see in reference \cite{PDG}).

The electromagnetic current $J^{\mu}$, is expressed in 
terms of the quark fields and the charge, i.e.,
~$J^{\mu}=\sum_{f} e_{f} \bar{q}_{f} 
\gamma_\mu q_{f}$, with $e_f$ and $q_f$ being, respectively, the charge and 
quark field of the flavor $f$ quark. 

\begin{figure}[htb]
	\centering
	\hspace*{3.80cm}
	\includegraphics[width=0.87\linewidth, trim={0.6cm 0.5cm 0.6cm 0.5cm}, clip]{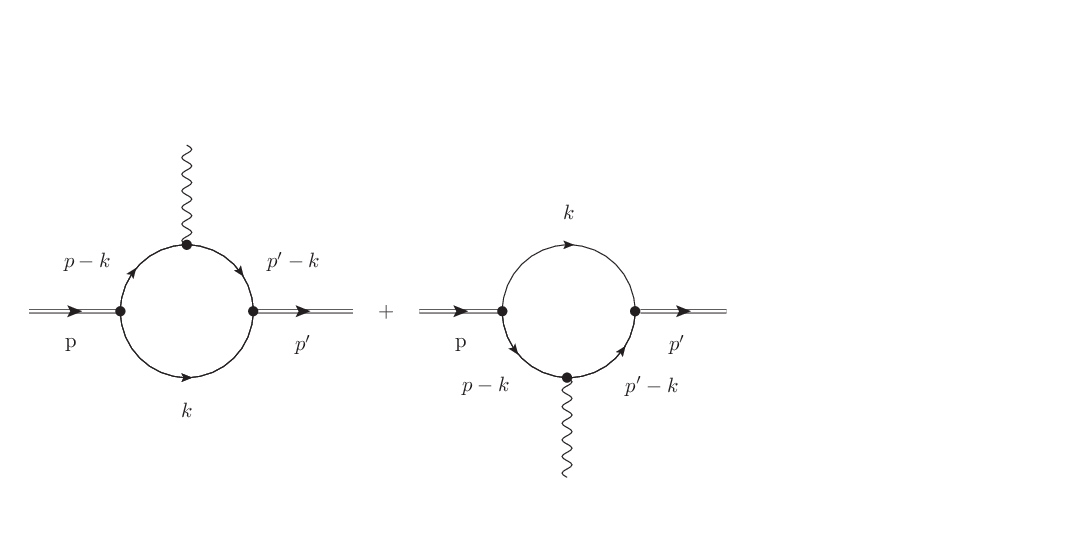}%
	\caption{Feynman triangle diagram with the valence contributions (left), 
		plus the non-valence contribution (right diagram). Due to mass symmetry, 
		only these diagrams are necessary to calculate the pion electromagnetic current.}
	\label{Fig1}
\end{figure}


The matrix element of the electromagnetic current is obtained
by the Mandelstam formula, with the
Feynman diagrams depicted in Fig.~\ref{Fig1}, 
\begin{eqnarray}
& & 
\Braket{ p';  M_{\pi^+} | j_1^\mu | p; M_{\pi^+} } 	  = -i N_c
\int \frac{d^4k }{(2 \pi)^4} g \Gamma (k,p') g \Gamma (k,p)
\nonumber	\\ & &
 \times \operatorname{Tr} \left[ 
\gamma^5  (\pi^{+})  S_q(k- p) 
\widehat{Q} \gamma^\mu S_q( k-p^{\prime})
\gamma^5 (\pi^{+})^{\dagger} S_q(k) 
\right]\hspace{-0.1cm} ,
\nonumber	\\
& & 
\Braket{ p';  M_{\pi^+} | j_2^\mu | p;  M_{\pi^+} } 
	  =-i N_c
\int \frac{d^4k }{(2 \pi)^4} g \Gamma(k,p') g \Gamma_M(k,p)
\nonumber	\\ & & 
\operatorname{Tr}
\left[
\gamma^5 (\pi^{+})^{\dagger} S_q(k+p') 
\widehat{Q} \gamma^\mu S_q( k+p)
\gamma^5   (\pi^{+}) S_q(k)
\right]\hspace{-0.1cm} ,
\label{jcurrent}
\end{eqnarray}
where the trace is performed over
the Dirac and flavor indices, and the total microscopic current is
	\begin{equation}
	\Braket{ p^\prime;  M_{\pi^+} | J^\mu| p;  M_{\pi^+} } = 
	\Braket{ p^\prime;M | j_1^\mu| p;M }  \hspace{-0.1cm} + \hspace{-0.1cm} 
	\Braket{ p^\prime;M | j_2^\mu| p;M } \, . 
	\label{fullcurrent}
	\end{equation}
Note,  the above expression  contains the
two diagrams shown in Fig.~\ref{Fig1}, and they are  written 
in flavor space and represent the photon being absorbed by each quark of the 
pion meson. 
	The contributions from flavor factors to the pion electromagnetic
	 current elements can be expressed separately as:
	\begin{equation}
	\text{Tr}[\dots] = 2 \left( \frac{1}{3} \Delta^\mu_{ u \bar d u} + 
	\frac{2}{3} \Delta^\mu_{\bar d u \bar d} \right)~,
	\label{delta12}
	\end{equation}
	where, for the $u$-quark on-shell, we have:
	\begin{equation}
	\Delta^\mu_{\bar d u \bar d} = \text{Tr}
	 \left[ \gamma^5 S^{\bar d} {q}(k- p)
	 \gamma^\mu S^{\bar d} {q}(k-p^{\prime})
	  S^u_{q}(k) \gamma^5 \right],
	\end{equation}
	
	and, analogously, for the $\bar{d}$-quark on-shell:
	
	\begin{equation}
	\Delta^\mu_{u\bar d u} = \text{Tr} \left[ \gamma^5 S^u_{q}(k- p)
	 \gamma^\mu S^u_{q}(k-p^{\prime}) S^{\bar d}_{q}(k) \gamma^5 \right].
	\end{equation}
	Assuming isospin symmetry for the pion, where the
	 masses of the $u$ and $d$ quarks are equal;
	  we set $\Delta^\mu_{u \bar{d} u} = \Delta^\mu_{\bar{d}u \bar d} =
	   \Delta^\mu$. Consequently, the expression simplifies to, 
	\begin{equation}
	\text{Tr}[\dots] = 2 \left( \frac{1}{3} \Delta^\mu + \frac{2}{3} \Delta^\mu \right) = 2 \Delta^\mu~.
	\end{equation}
	Thus, the electromagnetic current for the pion meson is given by the sum:
	\begin{equation}
	j^\mu_1 + j^\mu_2 = \langle p'|J^\mu|p \rangle = - 2 i N_c \int \frac{d^4k}{(2\pi)^4} \text{Tr} [ \Delta^\mu ] g \Gamma(k,p^{\prime}) g \Gamma(k,p) \ ,
	\label{jmu}
	\end{equation}
	
	In this work, we employ the light-front formalism 
	\cite{Brodsky1998,Dirac1949}. 
	The Light-front coordinates are given by, 
	 $a^+ = a^0 + a^3$, $a^- = a^0 - a^3$, and $\vec{a}_\perp = (a^1, a^2)$. The Dirac matrices in this formalism are represented as $\gamma^+ = \gamma^0 + \gamma^3$, $\gamma^- = \gamma^0 - \gamma^3$, and $\gamma_\perp = (\gamma^1, \gamma^2)$; and the metric is like,   
		$a \cdot b = \frac{a^+ b^- + a^- b^+}{2} - \vec{a}_\perp \cdot \vec{b}_\perp$.
		
		The kinematics for the four-momenta are defined as
		$p^\mu = (p_0, -\frac{q \cos \alpha}{2}, 0, -\frac{q \sin \alpha}{2})$ for
		the initial state and $p^{\prime \mu} = (p_0, \frac{q \cos \alpha}{2}, 0, \frac{q \sin \alpha}{2})$ for the final state. The transferred momentum is $q^\mu = (p^{\prime \mu} - p^\mu) = (0, q \cos \alpha, 0, \sin \alpha)$.
		With the Light-front coordinates, the momentum transfer is $q^\mu = (q^+, q^-, \vec{q}_\perp) = (q \sin \alpha, -q \sin \alpha, q \cos \alpha)$, with $q^2 = q^+ q^- - (\vec{q}_\perp)^2$. 
	
	The angle $\alpha$ allows for the choice of various reference frames, especially when $\alpha \neq 0$,
	because of contributions to electromagnetic current in addition to valence. This is important for illustrating the significance of the chosen reference frame.

Using the Mandelstam formula \cite{Mandelstam1955},
	 and performing the Dirac trace algebra, we evaluate the plus
	  component of the electromagnetic current $J^+$, where $\gamma^+ = \gamma^0 + \gamma^3$.
	   The elastic pion electromagnetic form factor, in terms of its full electromagnetic current, 
	   Eq.(\ref{fullcurrent}), is defined as 
	\begin{equation}
	F_\pi(q^2) = 
	\frac{\Braket{ p^\prime;  M_{\pi^+} | J^+| p;  M_{\pi^+}}}{2 p^+} ~.
	\end{equation}
 	As we know, when using the light-front approach, we have, in certain processes, a
 	 breaking of rotational invariance,
 	 not only for scalar mesons and pseudoscalar \cite{Naus1998,deMelo1999,Zhang2025,Bhamre2026}, but also 
 	 for spin-1 particles  \cite{deMelo1997,Karmanov1996,deMelo2026v1}.
 	  This breaking is accompanied by a dependence according to the adopted reference
 	   frame, and also, which component of the electromagnetic current is adopted to 
 	   extract the electromagnetic form factors~\cite{deMelo2002,deMelo2026}. 
 	    	 	In the present work, we explore the non-valence terms for spin-zero particles, in this case the pion.
 In the next section, we show our numerical results for the calculations of the observables for the pion,
  comparing them with experimental data, for both components of the 
  electromagnetic, $J^+$, and $J^-$.  Also, a detailed analysis of the contributions
 beyond valence is explored for both vertex models used.  
 We will see from the results presented that we have a pattern for the non-valence terms.
 
\section{Results}
\label{results}

This section presents the results of the 
present work. We examine the behavior of the zero-mode (or non-valence) contributions 
to the electromagnetic current and the form factor, calculated within the
 light-front approach using two different vertex models.

In Table (\ref{table1}), we present the results for the electromagnetic radius,
 the electroweak pion decay constant, and the Tarrach relation \cite{Tarrach1979}, 
 along with their
  dependence on the quark mass , calculated with the plus component of the 
  electromagnetic current. From this table, we can infer that a quark mass 
  of $m_q = 220~\text{MeV}$ yields a satisfactory result for both
 vertex models used here, when compared with the experimental data~\cite{PDG}.

\begin{table*}[htbp]
	\centering
	\caption{Values obtained for the electromagnetic radius of the pion, the pion electroweak decay
		constant, and the corresponding dimensionless Tarrach relation (TR)
		~\cite{Tarrach1979}, as a
		function of the quark mass ($m_q$) for
		the non-symmetric and symmetric model vertices.}
	\label{tab:combined_models}
	\begin{tabular}{c ccc c ccc}
		\hline
		\hline 
		\toprule
		& \multicolumn{3}{c}{Non-Symmetric Model} & & \multicolumn{3}{c}{Symmetric Model} \\
		\cmidrule(lr){2-4} \cmidrule(lr){6-8}
		$m_q$ [MeV] & $\sqrt{\langle r_\pi^2 \rangle}$ [fm] & $f_\pi$ [MeV] & TR & & $\sqrt{\langle r_\pi^2 \rangle}$ [fm] & $f_\pi$ [MeV] & TR \\
		\midrule
		\hline 
		180 & 0.795 & 81.46  & 0.327 & & 0.865 & 80.04  & 0.348 \\
		200 & 0.740 & 81.47  & 0.326 & & 0.796 & 86.40  & 0.349 \\
		220 & 0.696 & 92.54  & 0.326 & & 0.744 & 92.47  & 0.348 \\
		250 & 0.643 & 100.11 & 0.327 & & 0.681 & 101.19 & 0.349 \\
		280 & 0.601 & 107.14 & 0.328 & & 0.601 & 109.43 & 0.350 \\
		300 & 0.577 & 111.56 & 0.327 & & 0.577 & 114.60 & 0.352 \\
		320 & 0.555 & 115.78 & 0.325 & & 0.555 & 119.82 & 0.354 \\
		350 & 0.528 & 121.73 & 0.325 & & 0.528 & 127.23 & 0.358 \\
		380 & 0.505 & 127.32 & 0.326 & & 0.505 & 134.39 & 0.366 \\
		400 & 0.490 & 130.85 & 0.324 & & 0.490 & 138.95 & 0.372 \\
		500 & 0.433 & 146.51 & 0.322 & & 0.433 & 160.57 & 0.443 \\
		\midrule
		\hline 
		Exp.(PDG) & \ \ \ \ \ \ \ \  0.659$\pm$0.004 & \ \ \ \ \ \  \   92.21$\pm$0.141 &
		\ \  \ \  \ \  \ \ 0.314$\pm$0.004 & & -- & -- & -- \\
		\bottomrule
		\hline
		\hline
		\label{table1}
	\end{tabular}
\end{table*}

\begin{figure}[htb]
	\centering
	\includegraphics[width=0.47\linewidth]{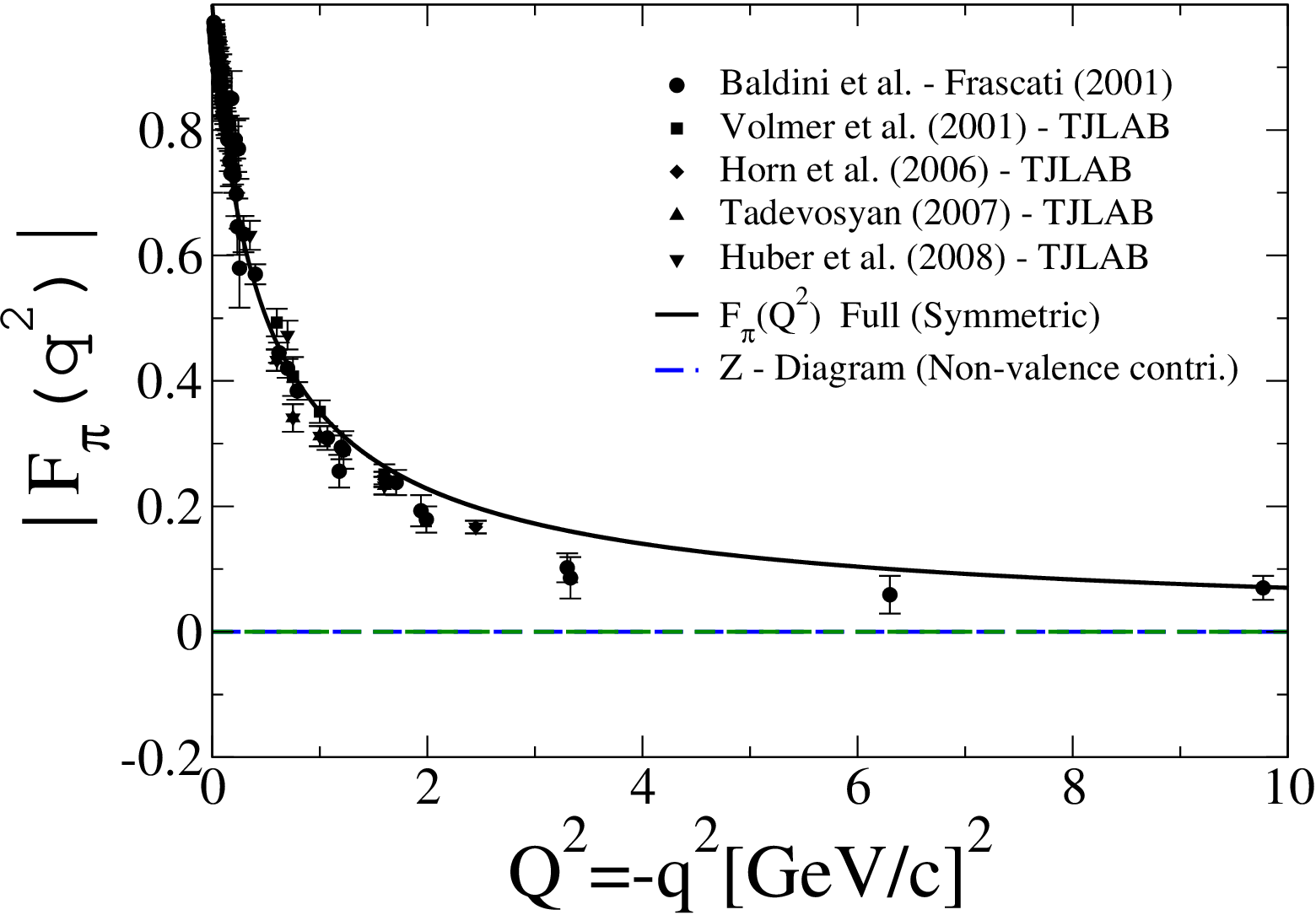}  
	\hspace{0.2cm}
	\includegraphics[width=0.47\linewidth] {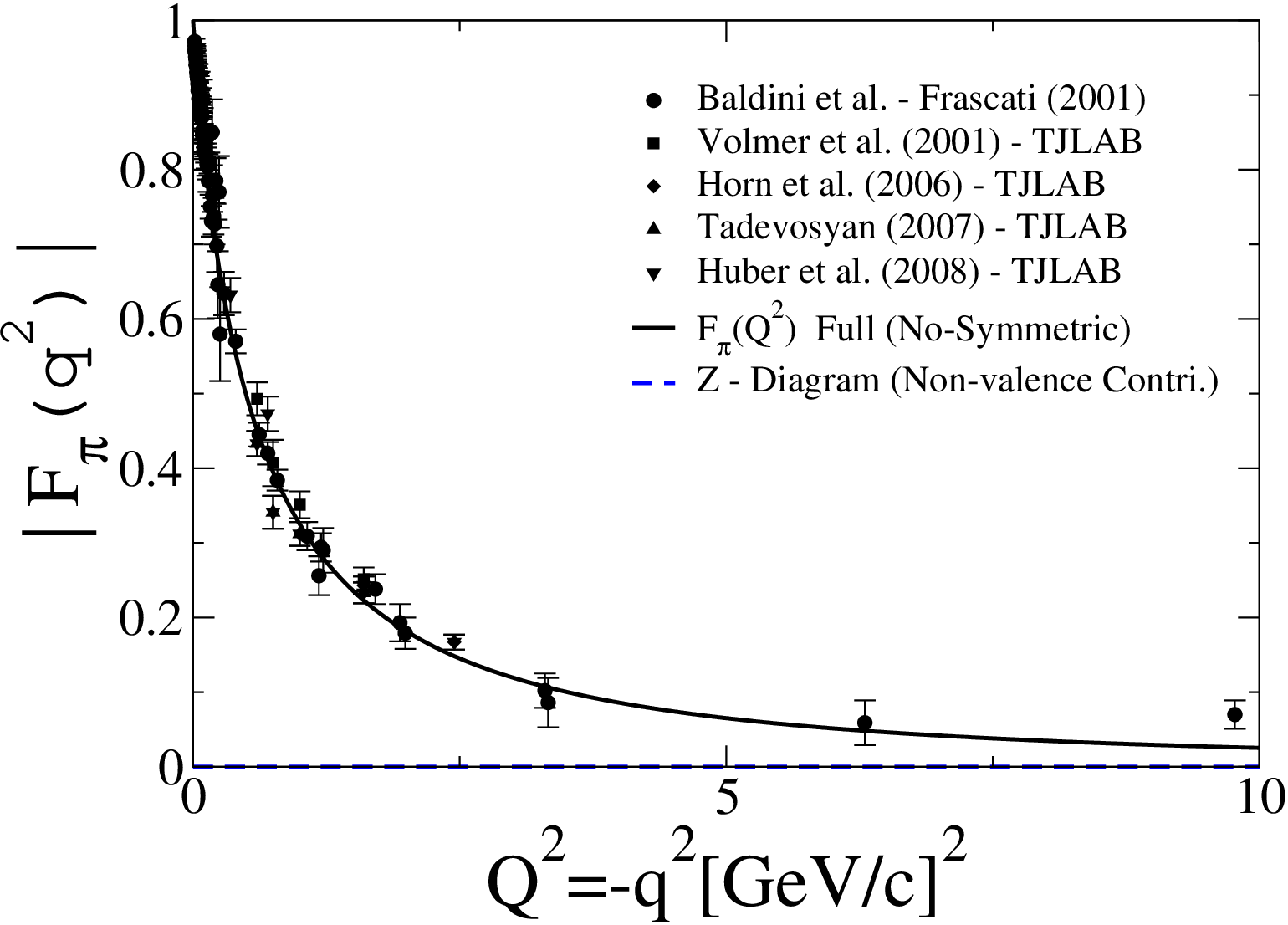}
	\caption{Full pion electromagnetic form factor for $\alpha=0$ degrees.
		In this case, we not have the non-valence contribution for the pion form factor, for 
		both models; compared with experimental data.
		For the symmetric model, the quarks masses are $220~MeV$
		and $M_R=600~MeV$; for the non-symmentric vertex case, quarks masses are also, $220~GeV$,
		but $M_R=1.0~GeV$.  Both vertex models, not have any non-valence contribution in the 
		frame with alpha zero, or $q^+=0$.}
	\label{Fig2}
\end{figure}

In Fig. \ref{Fig2}, 
we present the full electromagnetic form factor, calculated using the "plus" component 
of the electromagnetic current. Our results are compared with experimental data from Frascati 
and  JLab \cite{Baldini1999, Volmer2001, Horn2006, Tadevosyan2007, Huber2008}. 
In these calculations, we employ the frame where $\alpha=0$, corresponding 
to the $q^+=0$ component of the momentum transfer, which 
coincides with the Breit frame under the Drell-Yan condition.

This frame is advantageous as it leads to maximum suppression of the non-valence contributions,  
as illustrated in Fig. \ref{Fig2} (dashed green and blue line). 
For both  vertex models employed in this work, 
we find good agreement with the experimental data, following the figure.

\begin{figure}[htb]
	\centering
	\includegraphics[width=0.47\linewidth]{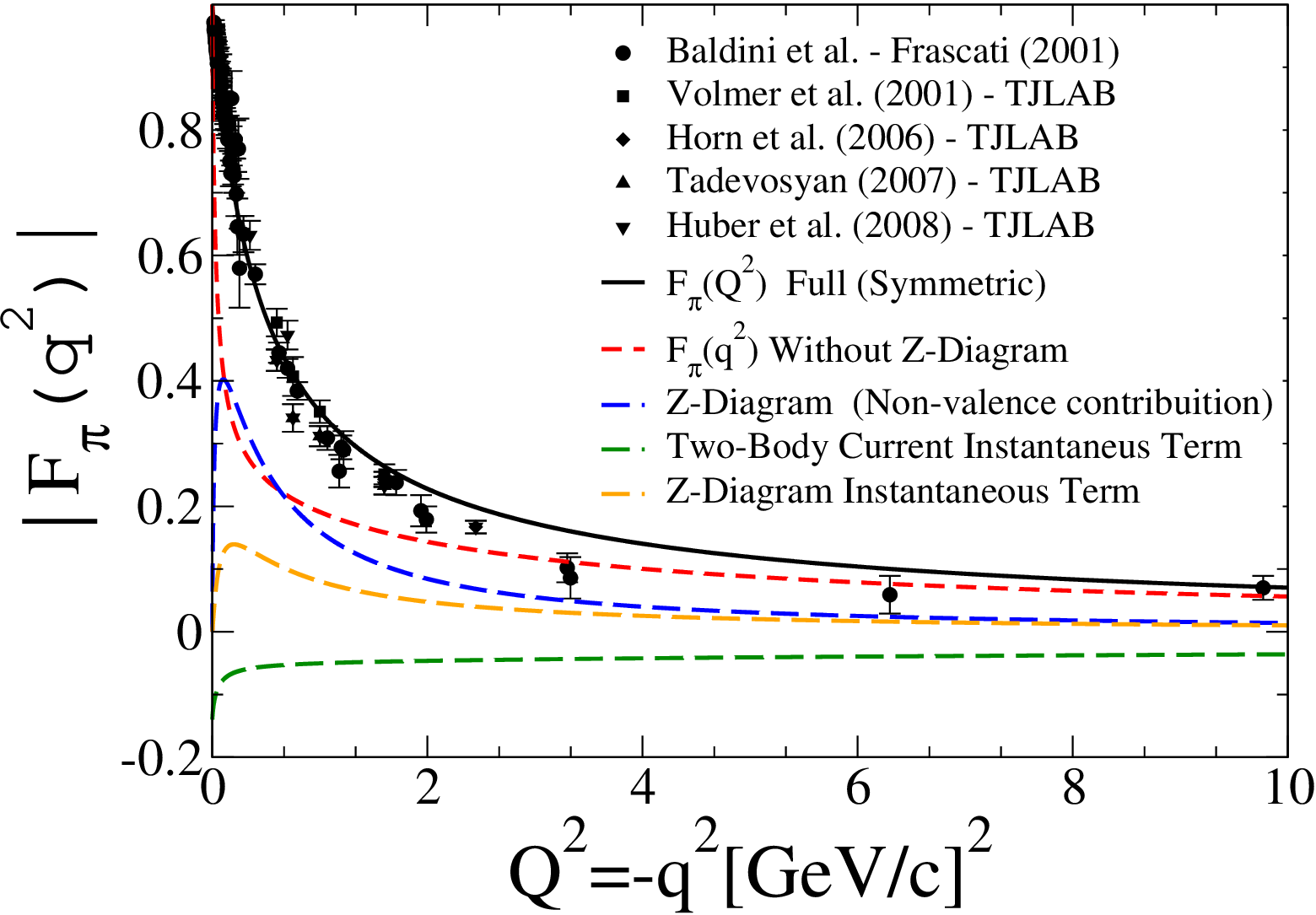}
	\hspace{0.20cm}
	\includegraphics[width=0.47\linewidth]{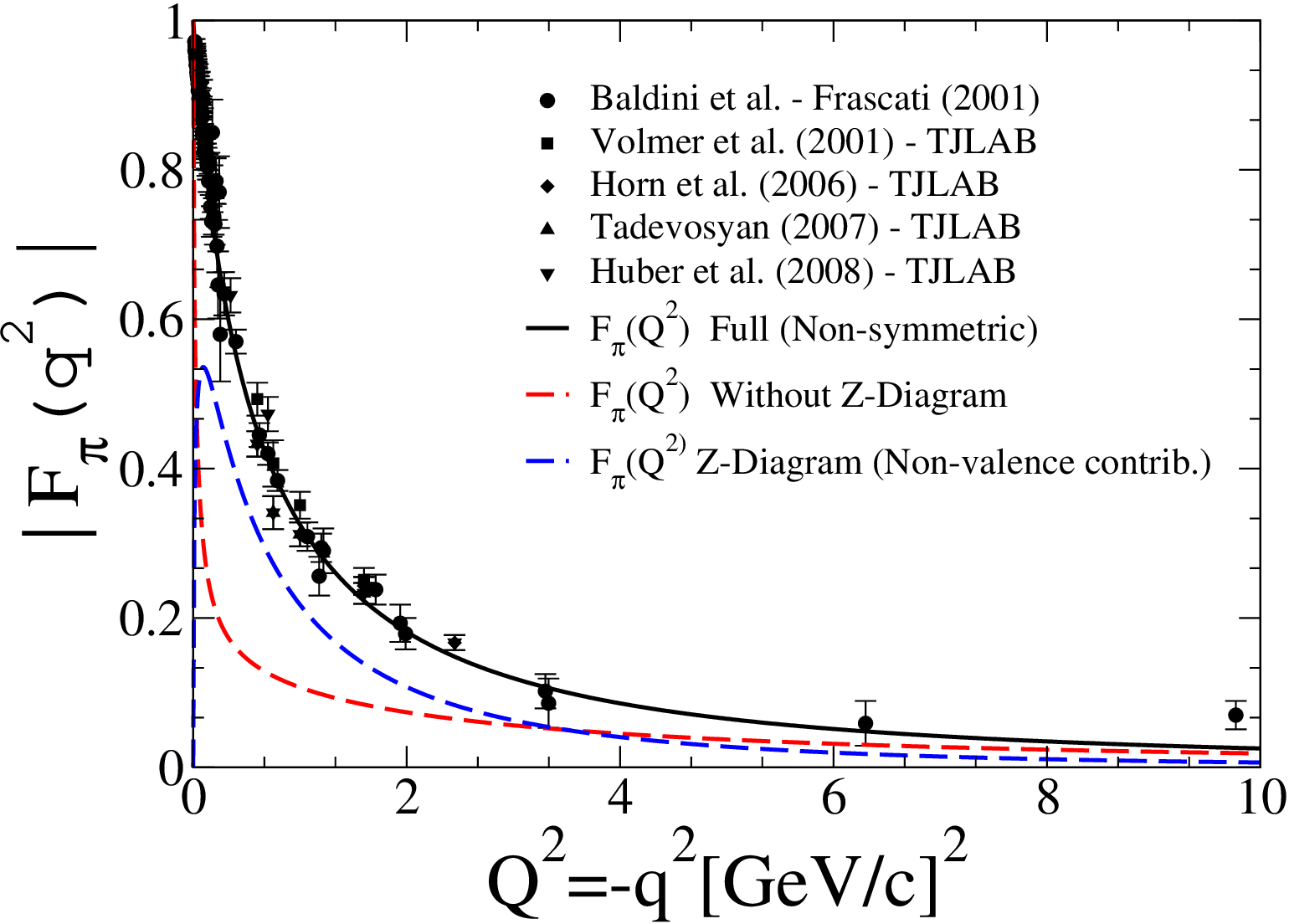}
	\caption{Same as Fig.\ref{Fig2}, but with alpha 45 degrees,
		which implies in $q^+\neq 0 $.} 
	\label{Fig3} 
\end{figure}

As previously explained, and as shown in Fig.\ref{Fig3}; using an angle other than zero ($\alpha \neq 0$),
we obtain a change of reference frame, now with a non-zero $q^+$, 
which leads to contributions to the electromagnetic current beyond the valence sector, ie., 
in addition to the valence terms. Once all contributions are taken into account, the final result
no longer depends on the reference frame~\cite{deMelo1997,deMelo1999,deMelo2002},
 as can be easily seen in the figure \ref{Fig3}.

\begin{figure}[htb]
	\centering
	\includegraphics[width=0.47\linewidth]{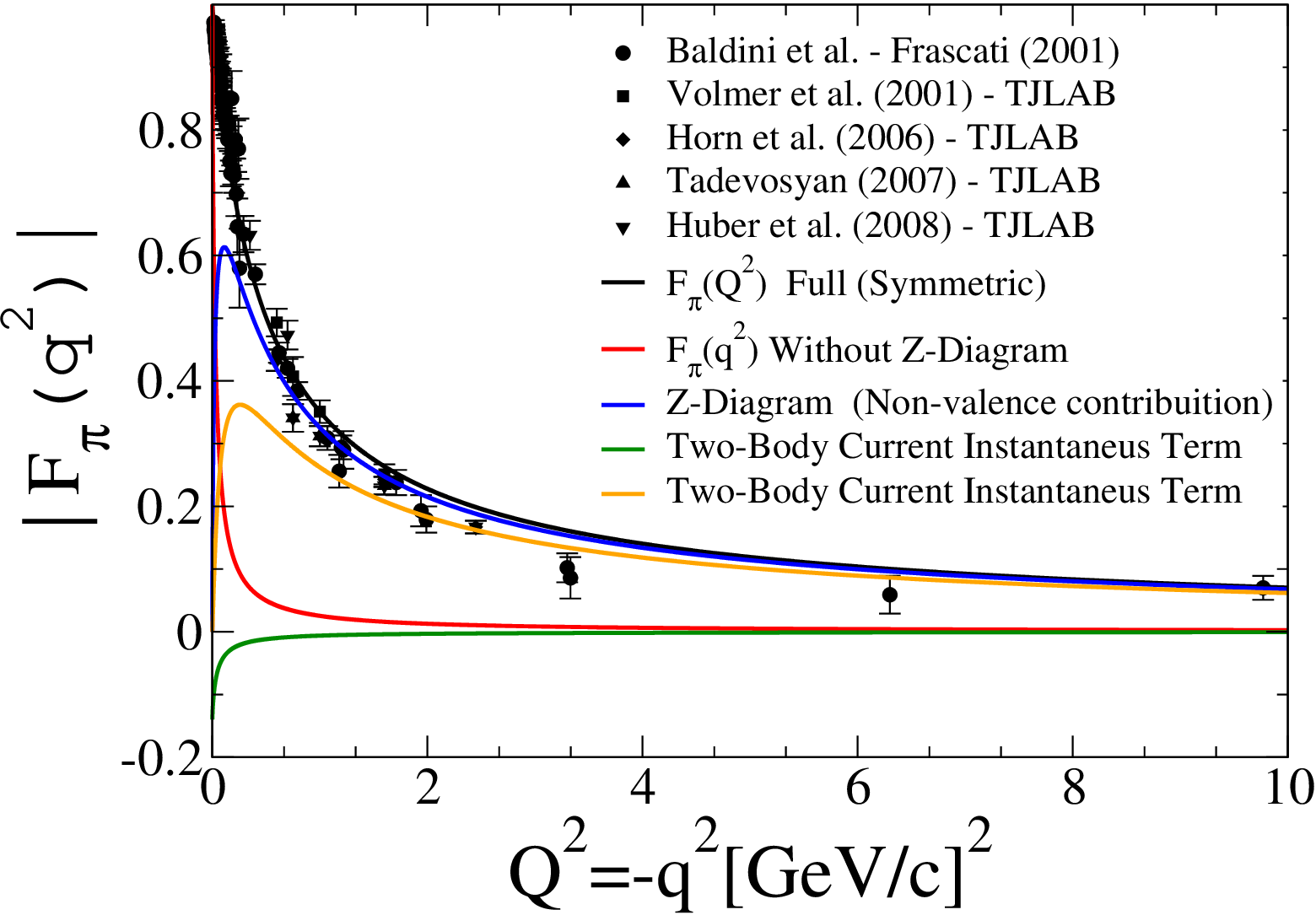}
	\hspace{0.20cm}
	\includegraphics[width=0.47\linewidth]{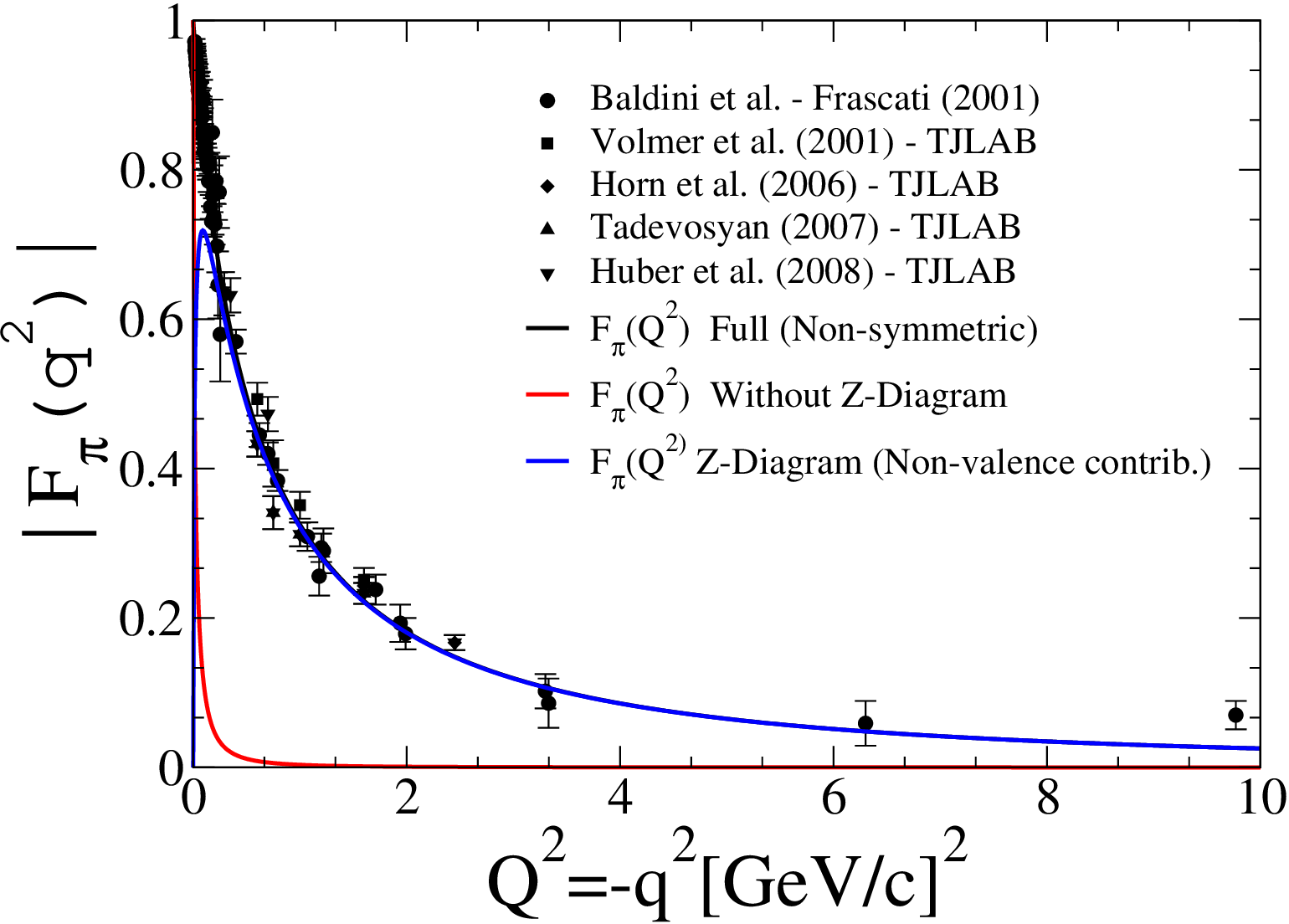}
	\caption{
		Pion electromagnetic form factor calculated in the 
		frame with alpha 90 degrees~($q^+ \neq 0$).}
	\label{fpifullnosialfa90}
\end{figure}

  In Figure \ref{fpifullnosialfa90}, where $\alpha = 90^\circ$, 
 one can observe the maximum contribution of terms other than the valence terms to the electromagnetic current; this substantially modifies the results for the pion electromagnetic form factor for both vertex models 
 used here.

  For the referential frames shown in the figures \ref{Fig3}, and 
  	 \ref{fpifullnosialfa90}
   ($\alpha \neq 0$, and $q^+ \neq 0$), after summing all contributions, 
   including the valence contributions to the electromagnetic current,
    which restores the model's covariance, 
    we find that the pion electromagnetic form factor reproduces the experimental data~\cite{Baldini1999,Baldini2000,Tadevosyan2007,Huber2008} quite well for the models employed~\cite{deMelo1999,deMelo2002,Moita2021}.

\begin{figure}[htb]
	\centering
	\includegraphics[width=0.47\linewidth]{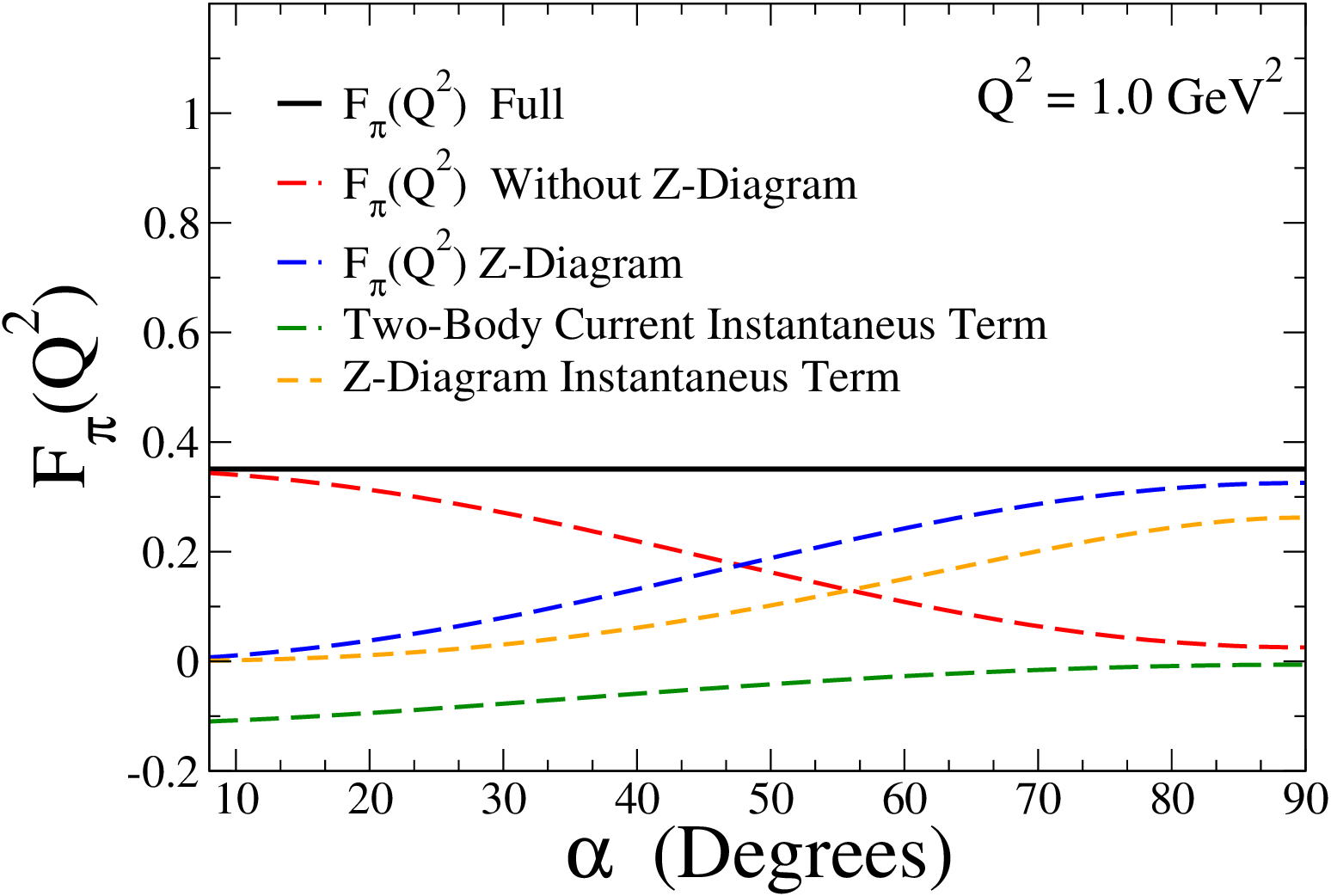}
	\hspace{.25cm}
	\includegraphics[width=0.47\linewidth]{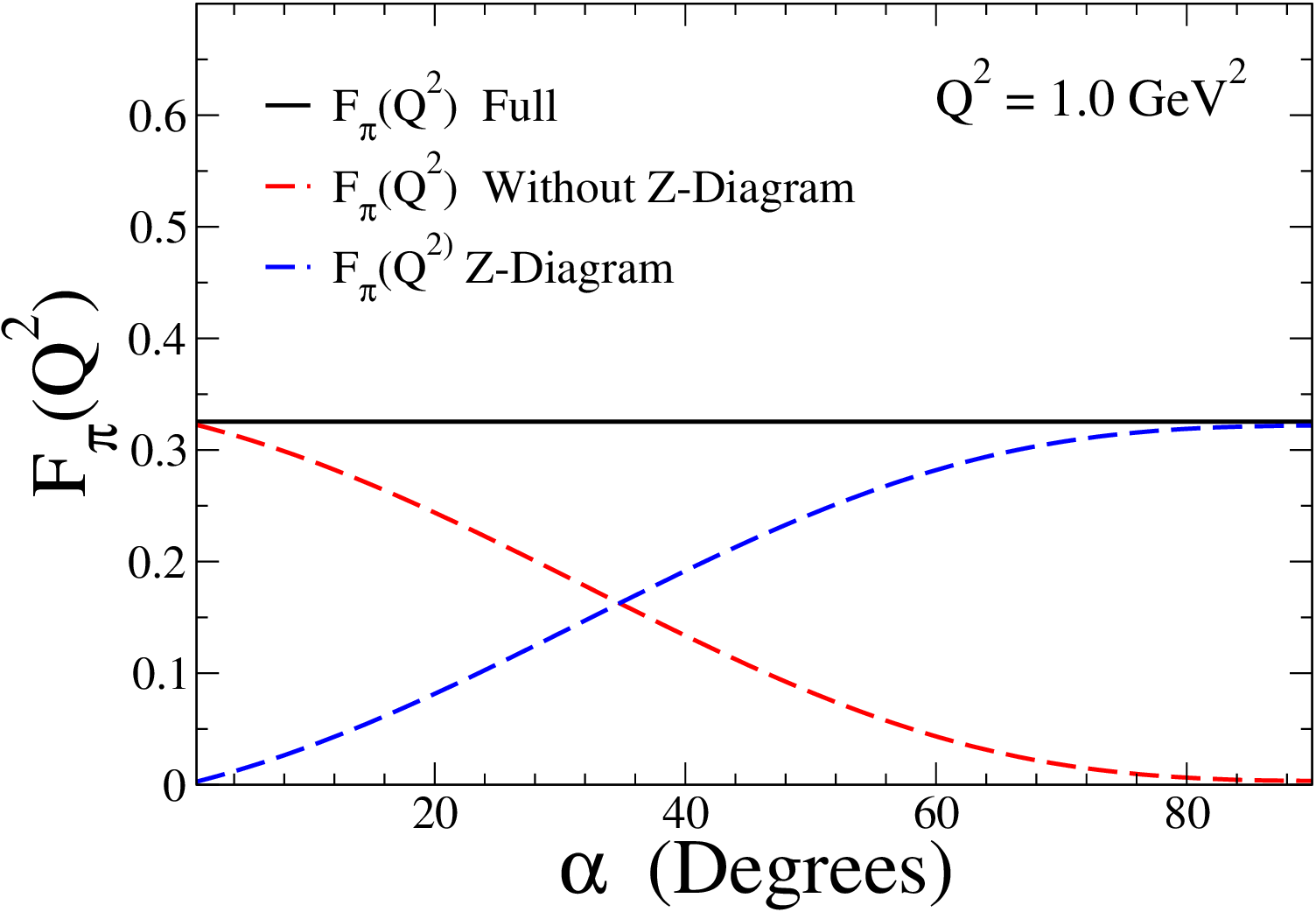} 
	\caption{All contribution for the pion electromagnetic form factor for two model 
	used in the present work, symmetric and non-symmetric vertex, with the 
momentum transfer fixed at $Q^2=1~GeV^2$. In the symmetric vertex case, 
the quarks masses are $0.220~GeV$ and $m_R=0.600~GeV$. 
For non-symmetric vertex, the masses are also 0.220~GeV, but 
$m_R=1.0~GeV$.}
	\label{fpivarq21}
\end{figure}

\begin{figure}
	\centering
	\includegraphics[width=0.47\linewidth]{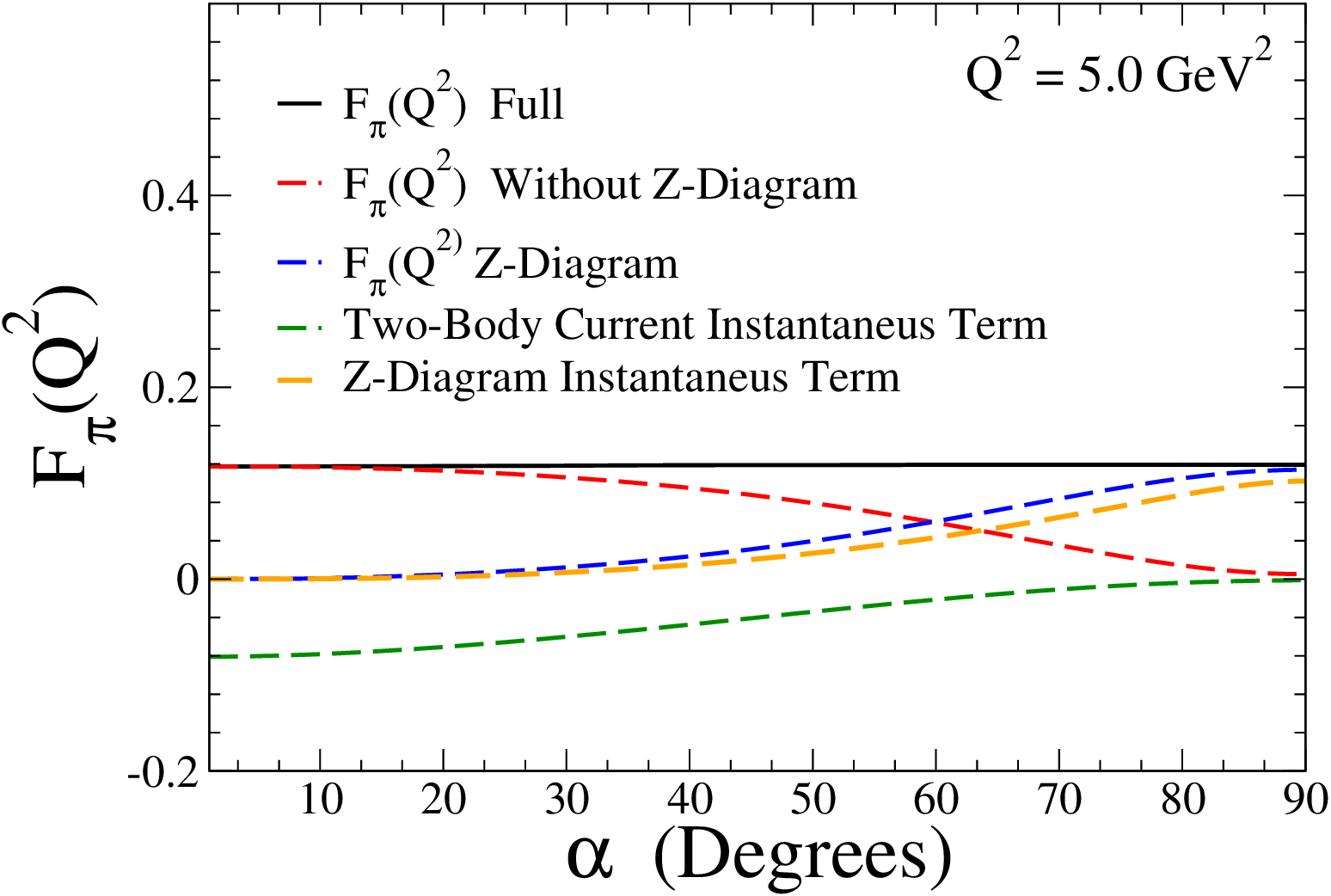}
	\hspace{0.25cm}
	\includegraphics[width=0.47\linewidth]{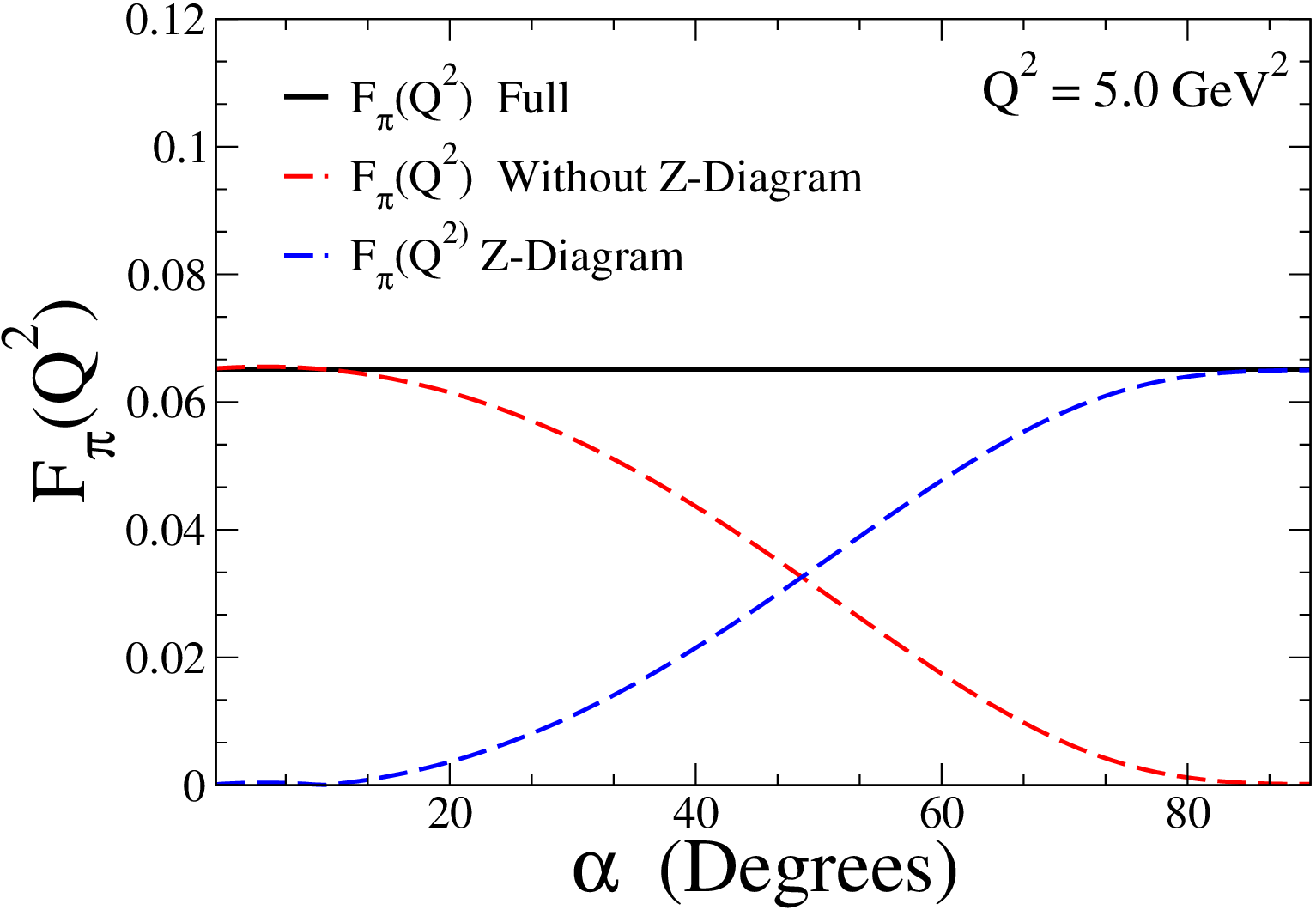}
	\caption{All contribution for the pion electromagnetic form factor for two model 
	used in the present work, symmetric and non-symmetric vertex, with the 
	momentum transfer fixed at $Q^2=5~GeV^2$. In the symmetric vertex case, 
	the quarks masses are $0.220~GeV$ and $m_R=0.600~GeV$. For non-symmetric vertex, the masses are also 0.220~GeV, but 
	$m_R=1.0~GeV$.}
	\label{fpivarnosyq25}
\end{figure}

 In the Figs. \ref{fpivarq21} and \ref{fpivarnosyq25}, the pion electromagnetic form factor is calculated for 
 both models of the vertex function used in the present work, with the momentum transfer, $q^2$
fixed at $1~GeV^2$ and $ 5~GeV^2$ respectively, and the angle $\alpha$ variation between the range 
$\alpha \in [0,90]$ degrees. 
These figures show significant frame dependence resulting from the variation of alpha. One can observe in the figures, 
particularly the one on the left, the various contributions to the electromagnetic form factor,
 in addition to the valence contribution. When all these contributions are summed,
  we obtain a constant value for the pion electromagnetic form factor; this is the expected outcome if the calculations are frame-independent, as they should be.

  Within the Light-Front Quantum Field Theory (LFQFT) framework, evaluating hadronic observables like the pion elastic electromagnetic form factor requires careful accounting of non-valence contributions, 
  commonly referred to as zero-modes or $Z$-diagrams~\cite{deMelo1999}. That contributions are essential for restoring rotational covariance,   which is explicitly broken when working with a truncated Fock-space expansion or 
  choosing specific components of the electromagnetic current
   (such as $J^+$ or, $J^-$)~\cite{deMelo2026,Deepesh2026}.     
   
   In the end, the covariance for the Light-front approach, is related with 
  the zero-modes contributions to the matrix elements of the electromagnetic 
  current~\cite{deMelo1997,deMelo1999}.
    
\begin{figure}[p]
	\centering
	\includegraphics[width=0.42\linewidth]{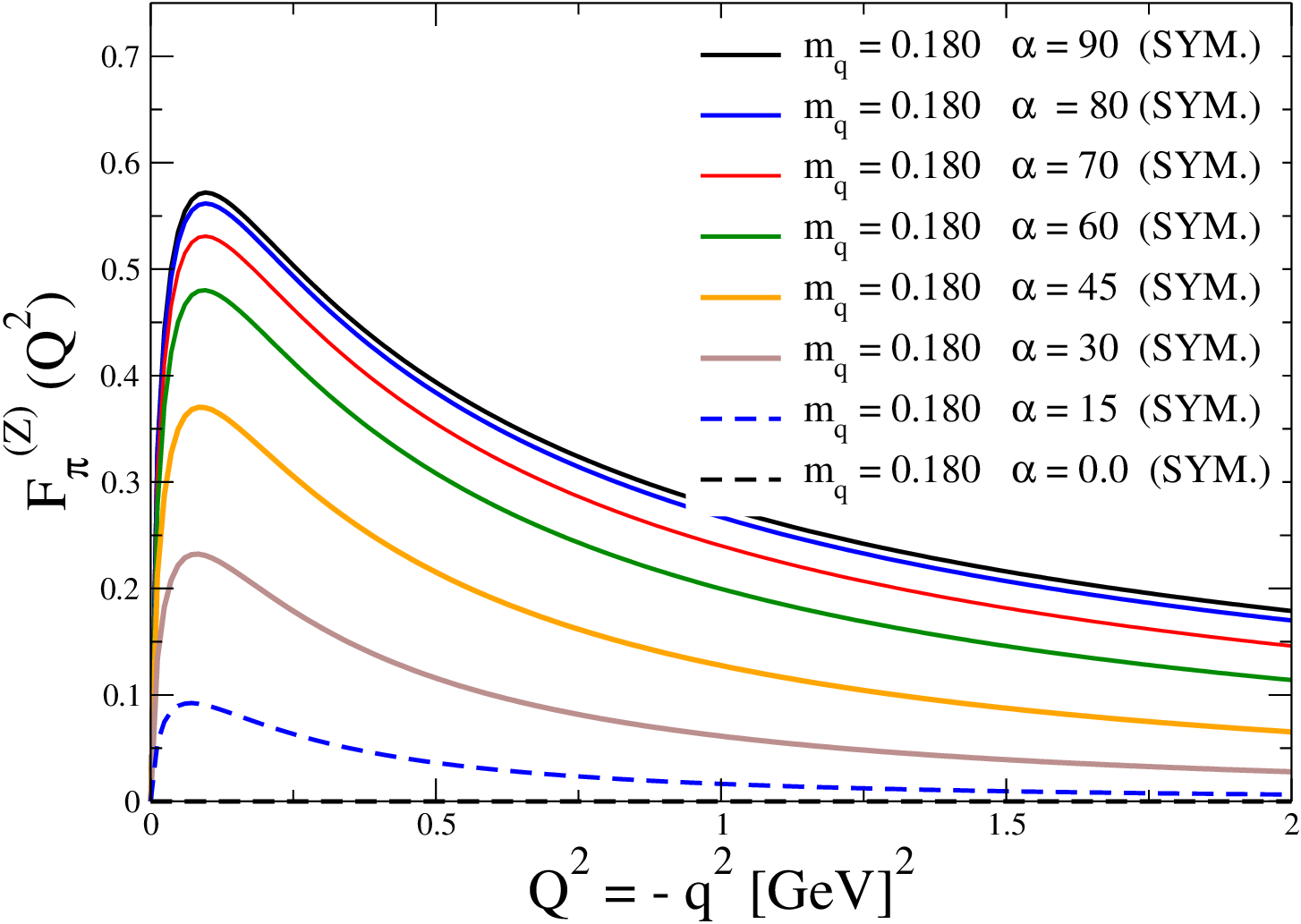}
	\hspace{0.2cm}
	\includegraphics[width=0.42\linewidth]{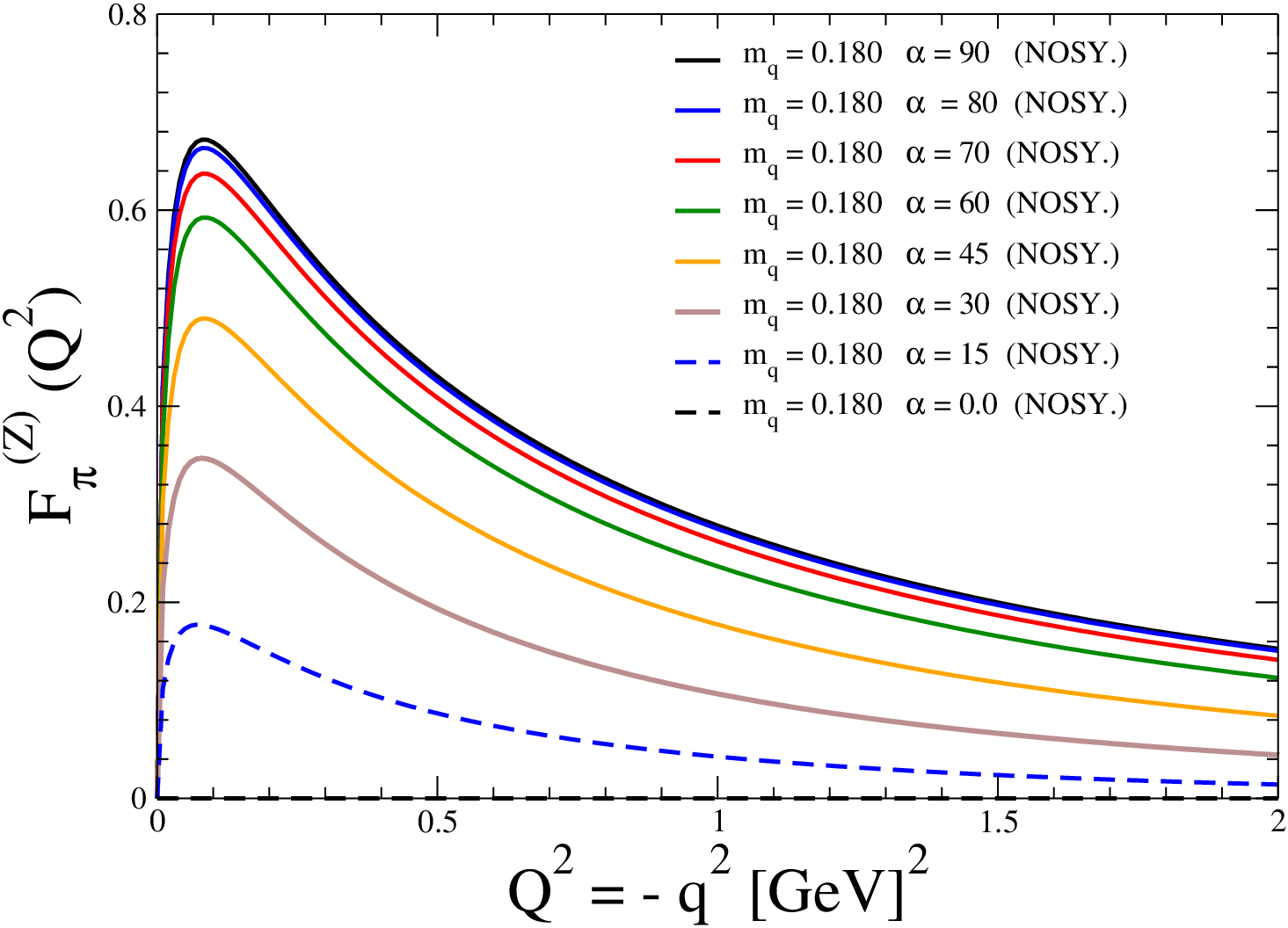}
	
	\vspace{0.3cm}
	
	\includegraphics[width=0.42\linewidth]{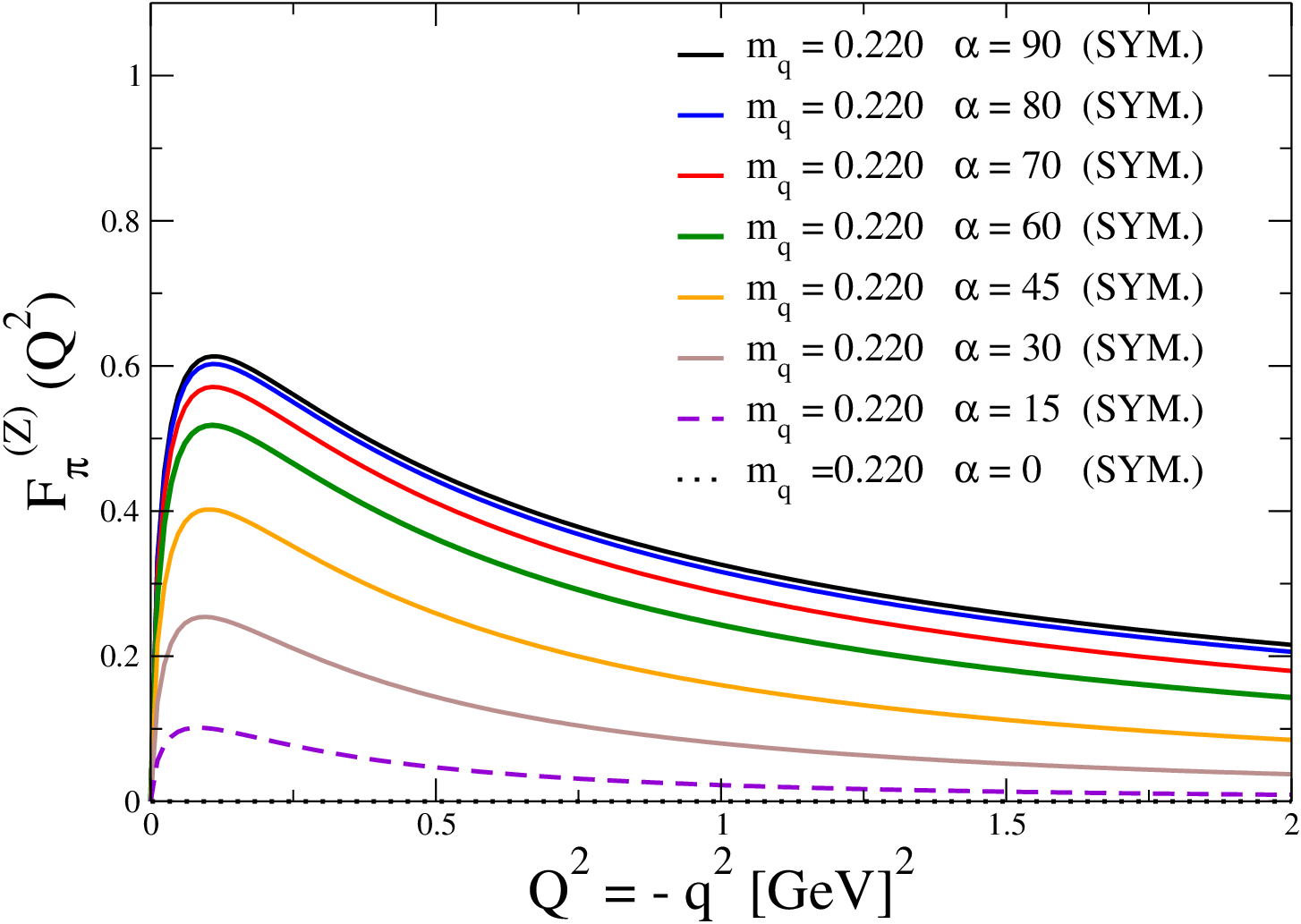}
	\hspace{0.2cm}
	\includegraphics[width=0.42\linewidth]{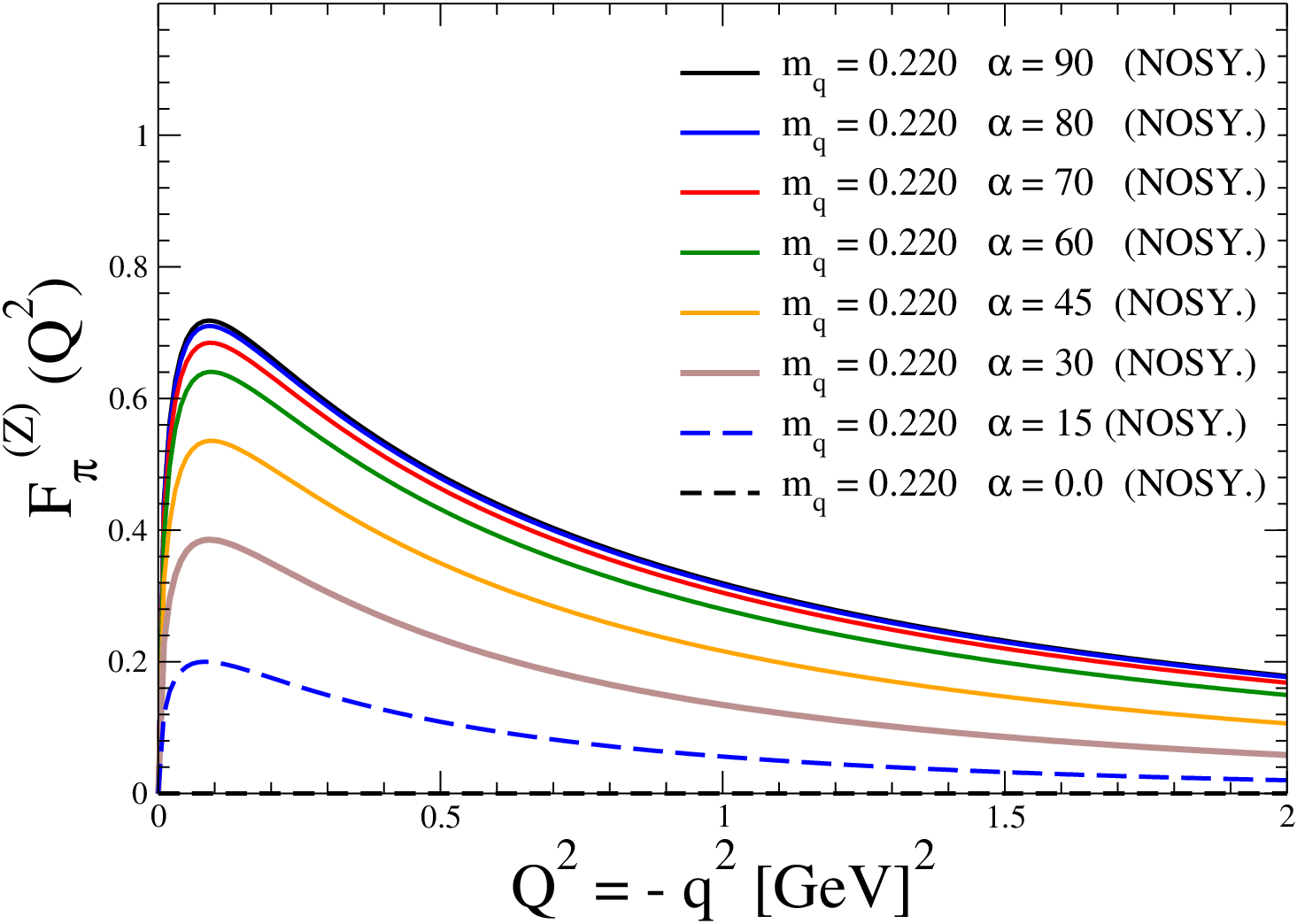}
	
	\caption{
		Z-diagram contribution to the quark masses.
		The upper row presents the results for the quark mass,
		$m_q = 0.180~\text{[GeV]}$, comparing the symmetric (left panel)
		and non-symmetric (right panel) vertex models. The regulator masses
		used are $m_R = 0.600~\text{[GeV]}$ for the symmetric model
		and $m_R = 1.0~\text{[GeV]}$ for the non-symmetric model.
		The lower row shows the corresponding results for a higher quark mass, $m_q = 0.220~\text{[GeV]}$,
		maintaining the same vertex models (symmetric on the left, non-symmetric on the right).
		Note that the maximum zero-modes contribution occurs at the same $Q^2$ for both models in each case.
		The non-valence (zero modes) contribution vanishes when $q^+ = 0$. In these figures,
		this situation corresponds to $\alpha = 0^\circ$.
	}
	\label{fig7}
\end{figure}

\begin{figure}
	\centering
	\includegraphics[width=0.45\linewidth]{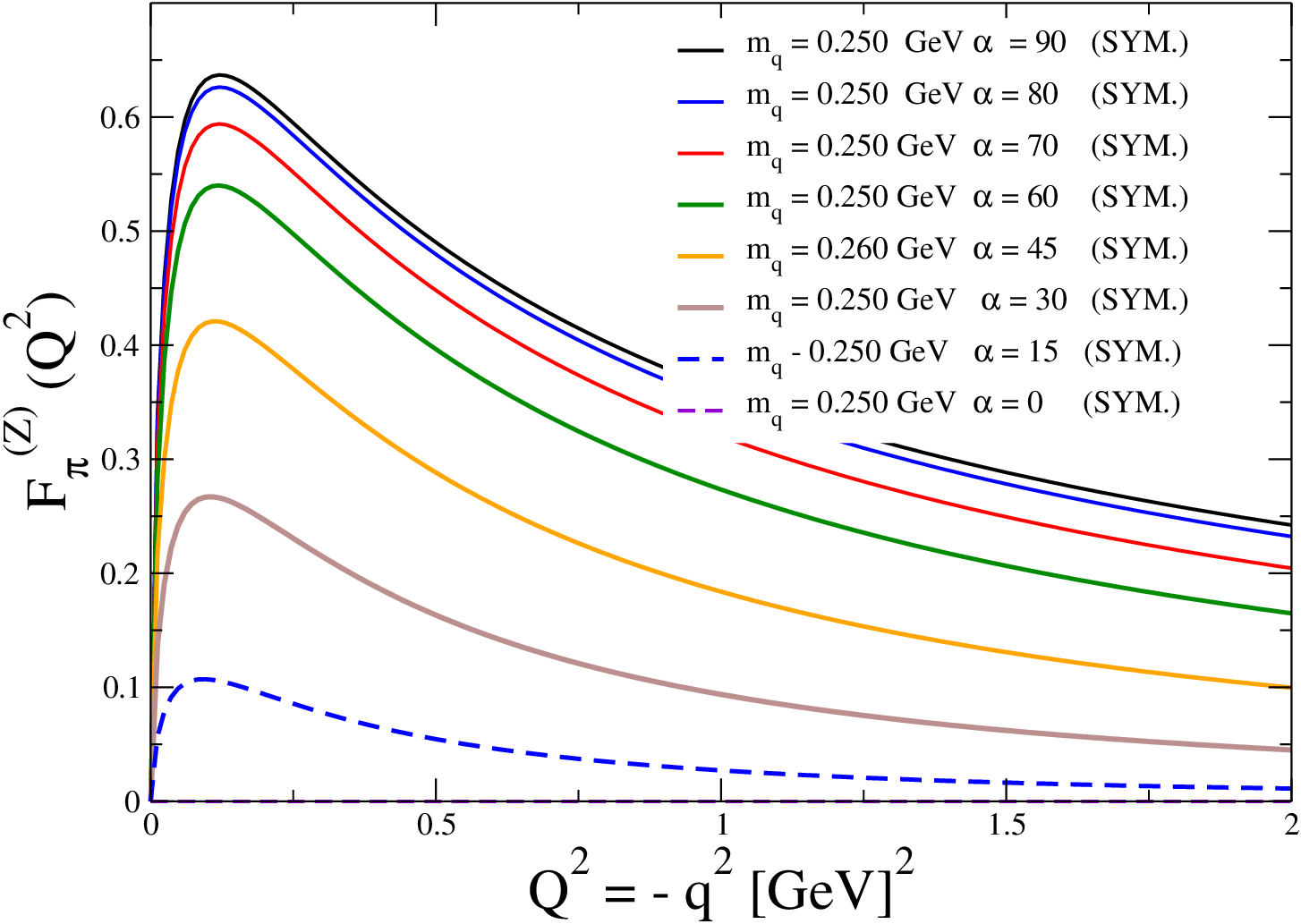}
	  \vspace{0.30cm}
	  	\includegraphics[width=0.45\linewidth]{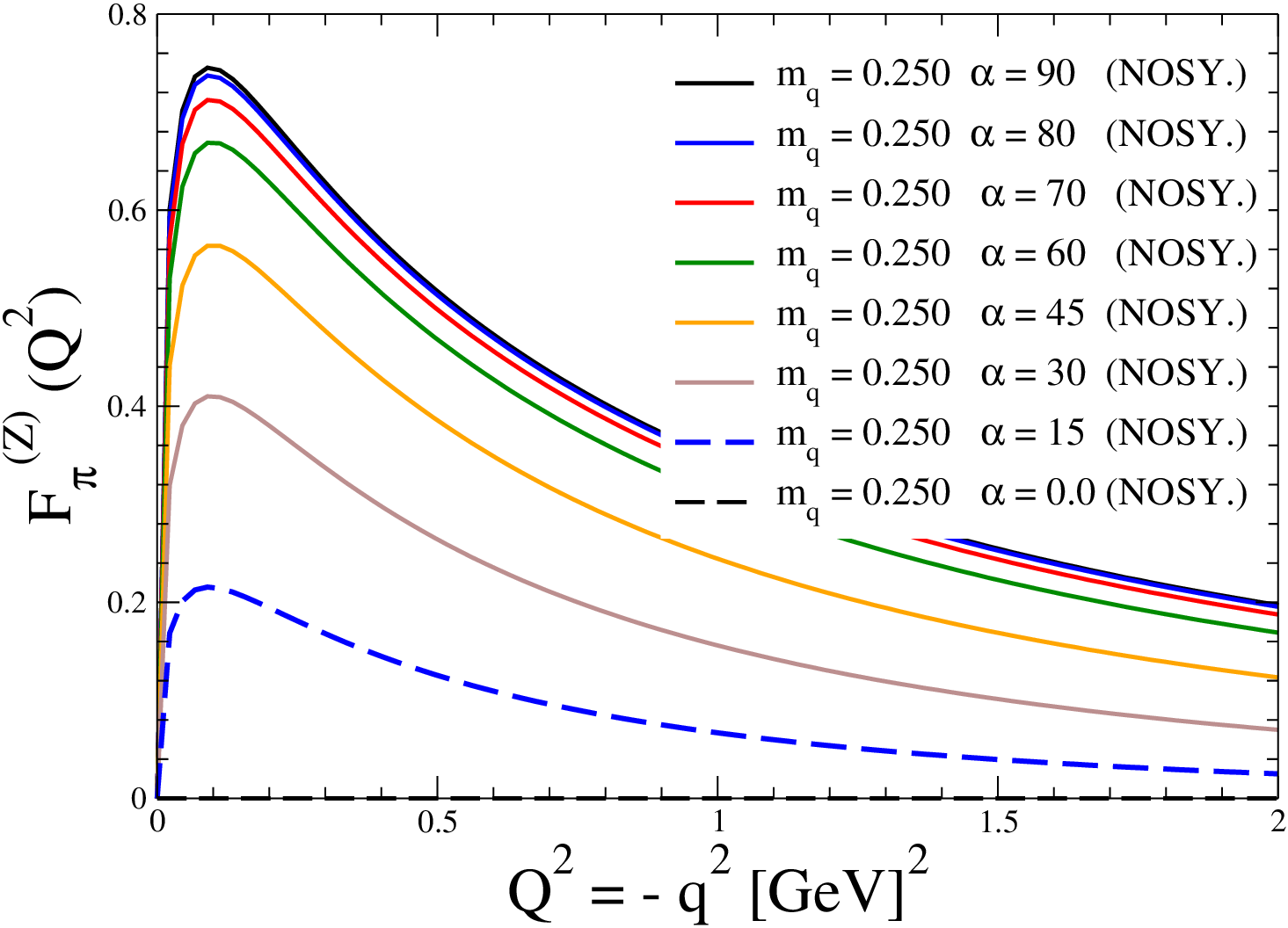}

 \vspace{0.55cm}

	\includegraphics[width=0.45\linewidth]{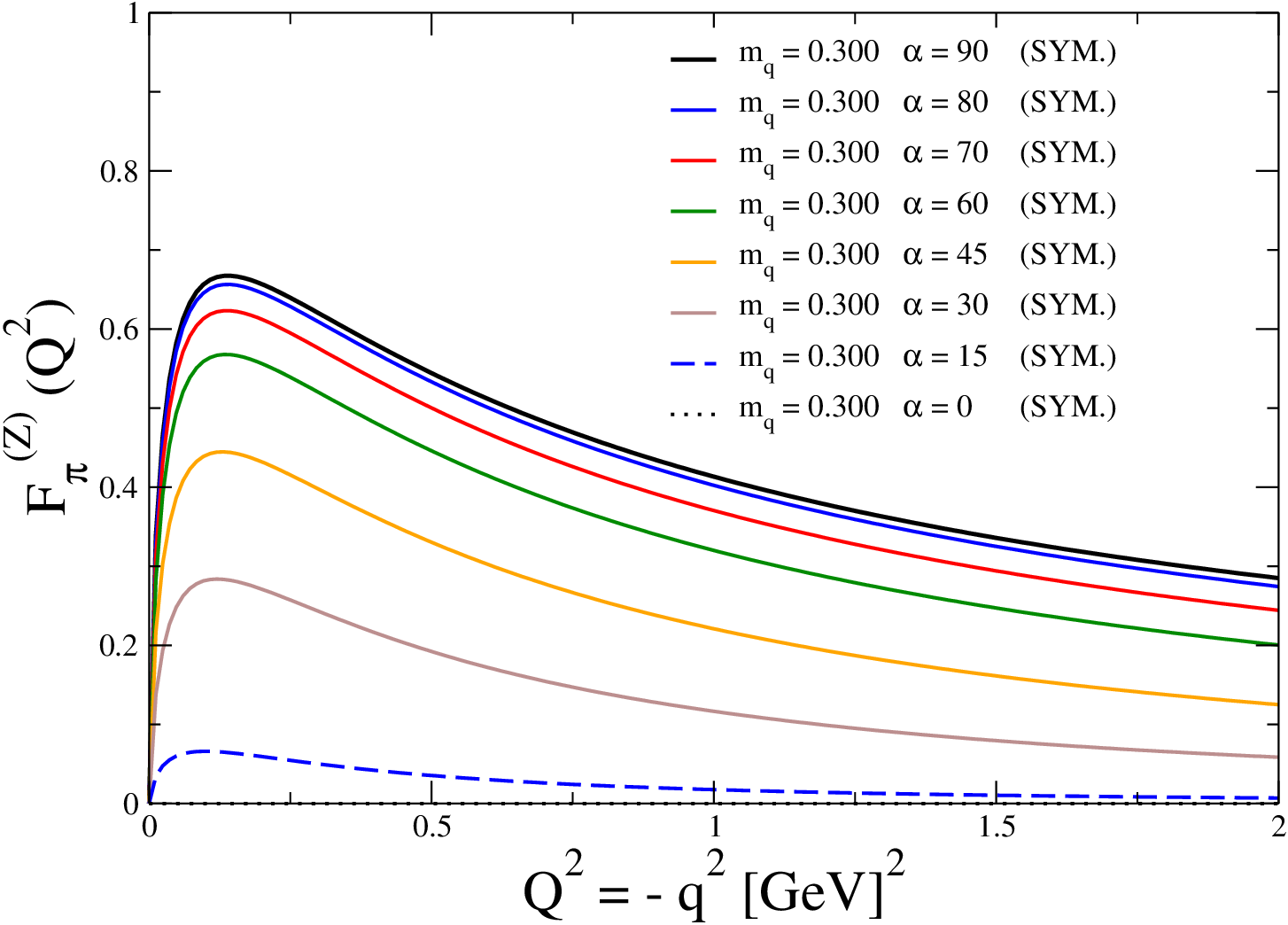}
\vspace{0.30cm}
\includegraphics[width=0.45\linewidth]{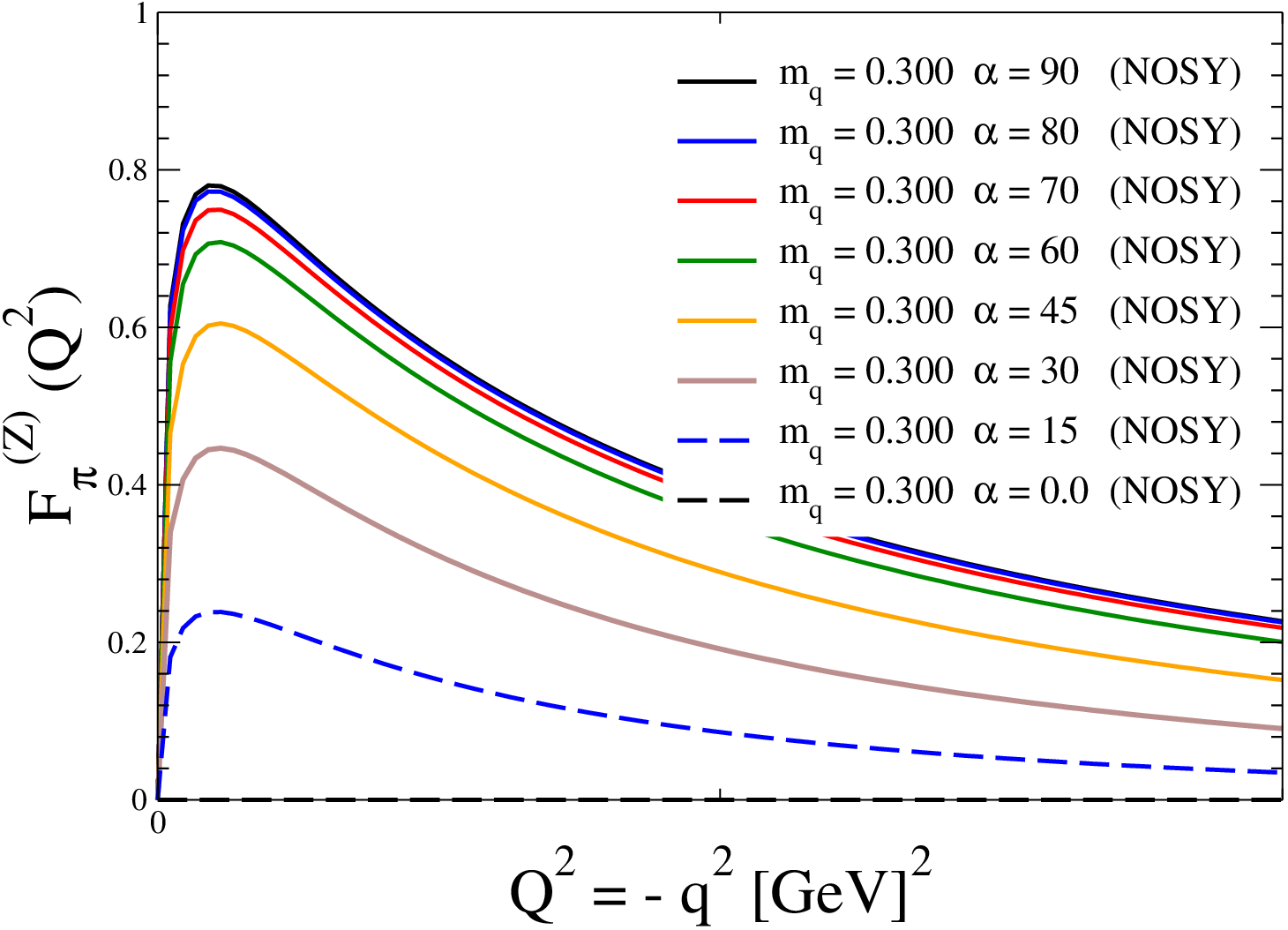}
	\caption{Z-diagramm contribution, for the quark masses 0.250 [GeV] (upper), and 
		$m_q$=0.300 [GeV]. Label the same of Fig. \ref{fig7}. 
		The figure show, the maximum contribution from zero modes appear in the 
		same momentum transfer.
}	 
	\label{fig8}
\end{figure}

\begin{figure}
	\centering
	\includegraphics[width=0.45\linewidth]{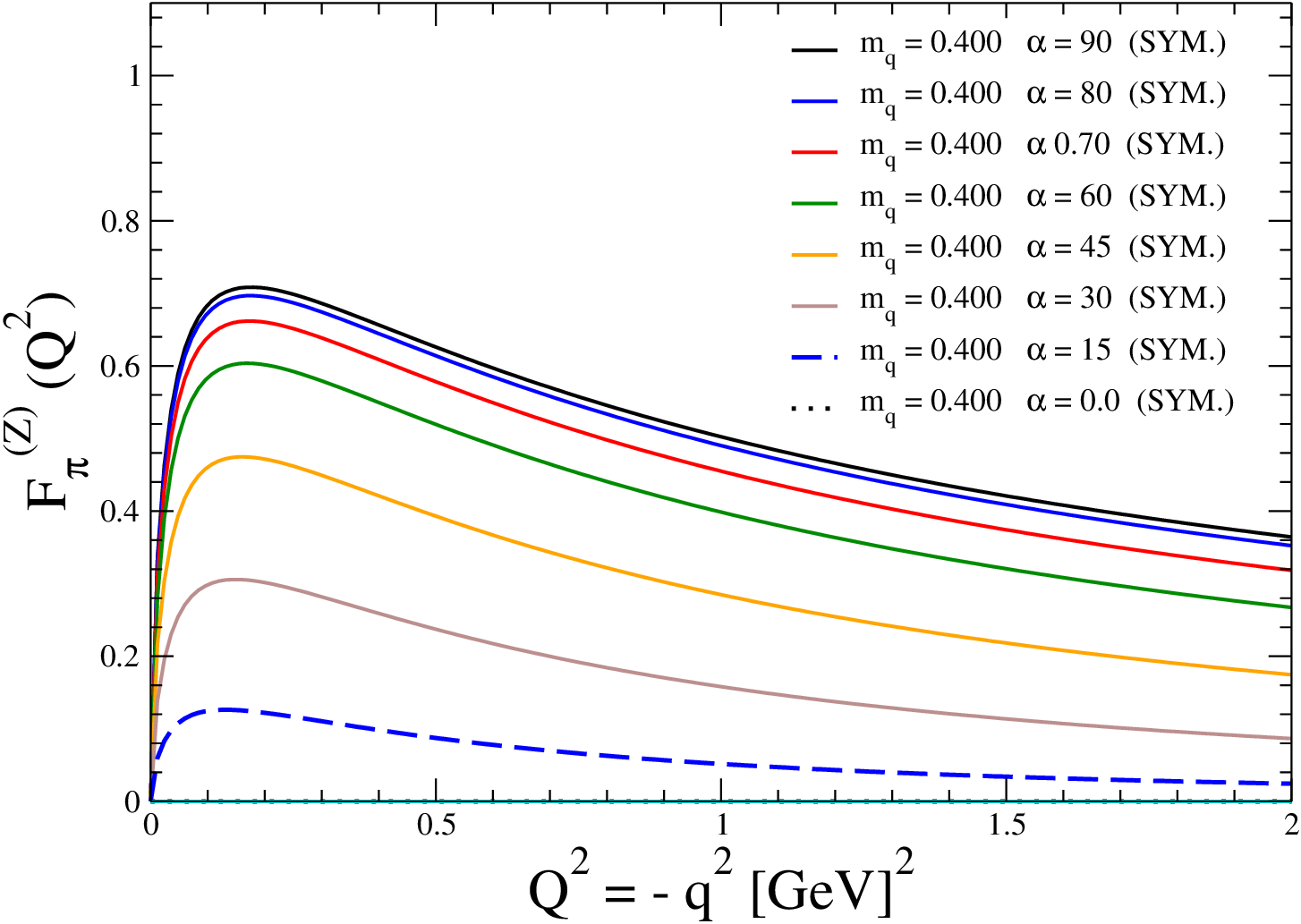}
	 \vspace{0.30cm}
		\includegraphics[width=0.45\linewidth]{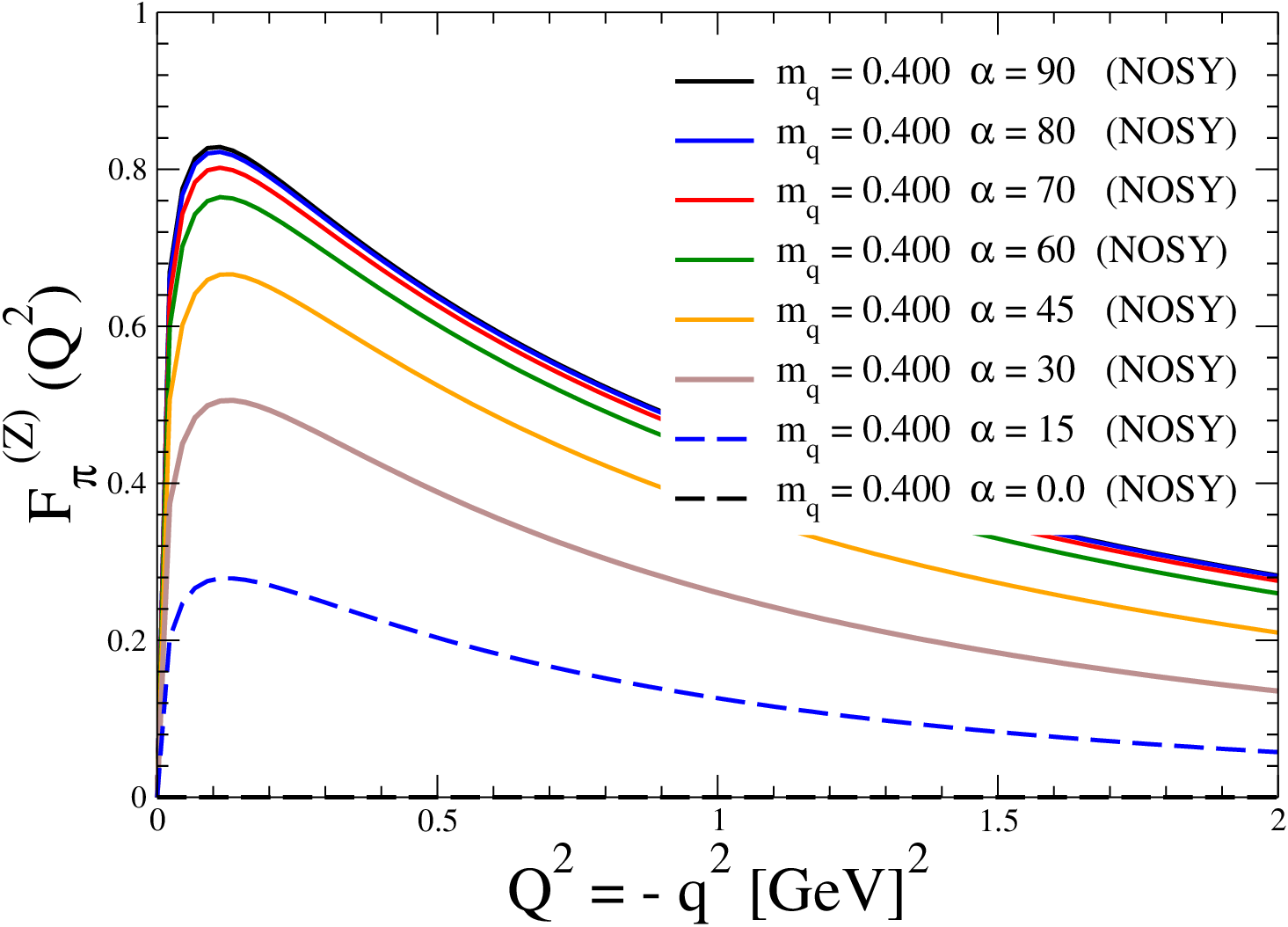}
		\caption{Z-diagramm contributions, for the quark masses, $0.400~GeV$. 
		The models vertex utilized in the present work is 
		from  the Refs.~\cite{deMelo2002,deMelo1999}. The maximum zero modes contribution,  
		happens, indpendente of the frame used, in the same $Q^2$ for both models.
	}
	\label{fig9}
\end{figure}

   With the figures \ref{fig7},\ref{fig8},  and \ref{fig9}, the zero modes 
  contributions, or Z-diagram contributions  for the pion elastic electromagneticc form factor 
  as function of the constituent quark masses are explored  for both vertex models.

  In the Fig.\ref{fig7}, the results for quark masses of
  $m_q = 0.180~\text{[GeV]}$ and $m_q=0.220~ GeV$; comparing the symmetric and non-symmetric 
  vertex models are presented.

  The regulator masses
  used are $m_R = 0.600$~GeV for the symmetric model
  and $m_R = 1.0$~GeV for the non-symmetric model. The calculations are done with the 
  alpha angle variantion~($0^\circ \leq \alpha \leq 180^\circ$); that means,
   a variantion in the referential frame. 
   
   We can see, in the figures, the maximum zero-mode contribution occurs at the 
   same $Q^2$ for both models in each case.
  The non-valence (zero-mode) contribution vanishes when $q^+ = 0$, which in these figures
  is consistent with   to $\alpha = 0^\circ$.

  As observed, the zero-mode contribution
  $F_\pi^{(Z)}(Q^2)$ exhibits a pronounced peak at a specific momentum transfer scale,
   denoted as $Q^2 = q_{Z,\text{max}}^2$. A key feature highlighted by these results is
    that the maximum of the zero-mode contribution always occurs at the exact 
    same $Q^2$ value, regardless of the choice of the reference frame angle $\alpha$
     (for $\alpha \neq 0^\circ$). 
     
     This means that, for a given constituent quark mass, the maximum
    contributions to the zero-modes of the electromagnetic current elements occurs
     at the same $Q^2$ transfer momentum, even when the reference frame is changed; 
     see the Figs. \ref{fig8}, and \ref{fig9} for masses of the quarks with the values $0.250, 0.300~GeV$ and $0.400~GeV$.

  
  In Fig.~\ref{fig7}, the contributions to the elastic electromagnetic form factor of the pion, ~$F_\pi(Q^2)$, due to zero modes or non-valence terms, are presented as a function of the variation of the angle $\alpha$, chosen between $0^\circ$ and $90^\circ$; which means that the reference frame is varying for each chosen mass of the constituent quarks. In the corresponding figure, we present the quark masses as being $180\text{ MeV}$ and $220\text{ MeV}$. We can see that the maximum contribution to the non-valence terms occurs at the same value of the transferred momentum ~$q_{z,\text{max}}^2$, for all angles $\alpha$ presented in this figure,
   for both the symmetric vertex model and the non-symmetric vertex model.

  In Figs.~\ref{fig8} and \ref{fig9}, the same strategy of varying the angle $\alpha$, with the consequent variation of the reference frame, is adopted with different quark masses; in Fig.~\ref{fig8}, the masses are equal to $250\text{ MeV}$ and $300\text{ MeV}$, respectively, whereas for Fig.~\ref{fig9} the quark masses are equal to $400\text{ MeV}$. The calculations are performed for both vertex models.
  
  We can clearly see a pattern in the maximum transferred momentum ~$q^2_{Z,\text{max}}$, which varies with the quark masses, but this maximum is independent of the reference frame, since the figures show $\alpha \in [0, 90]^\circ$ for various masses, always occurring at the same value.

\begin{figure}
	\centering
	\includegraphics[width=0.45\linewidth]{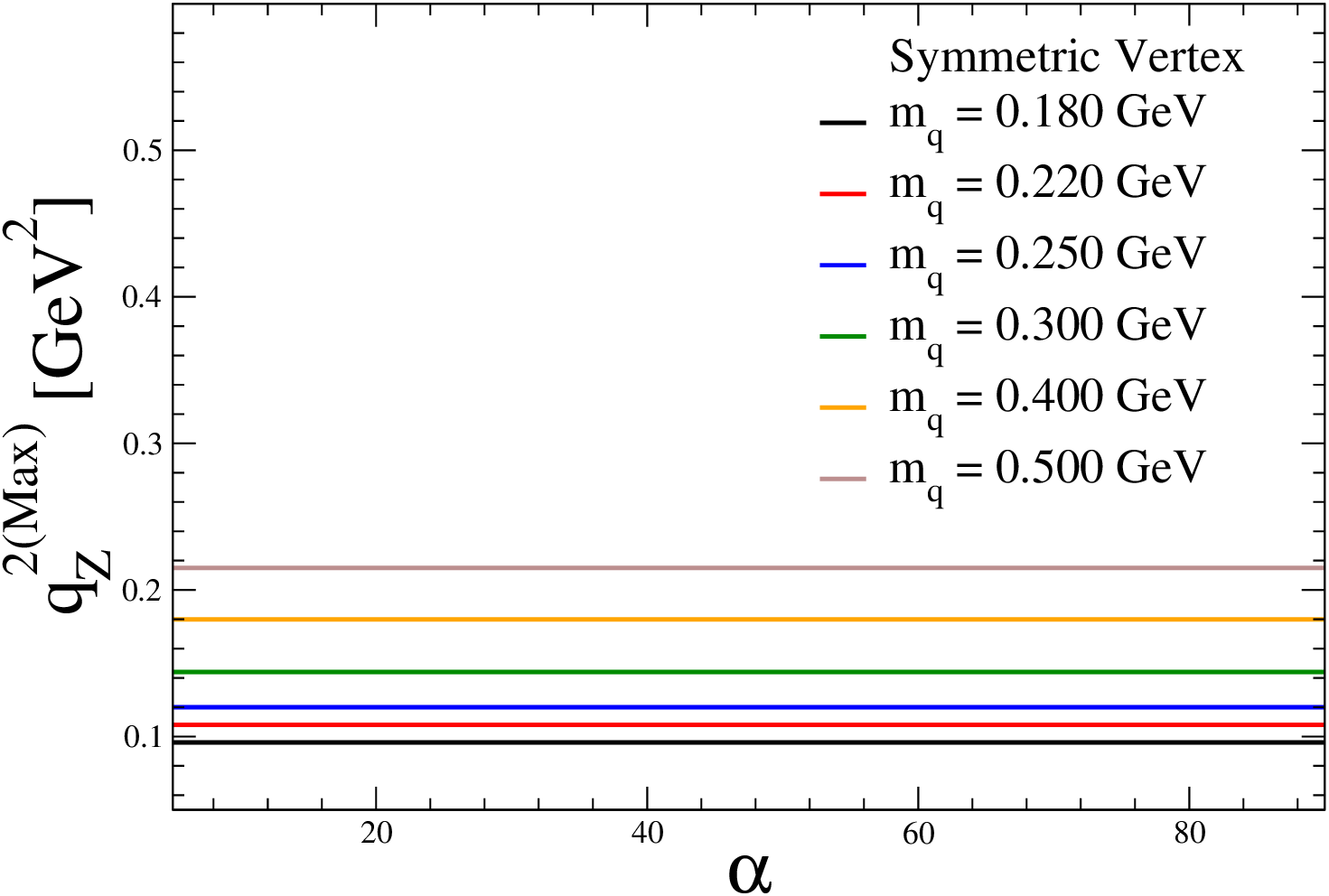}
{\tiny .}	\vspace{0.50cm}
	\includegraphics[width=0.45\linewidth]{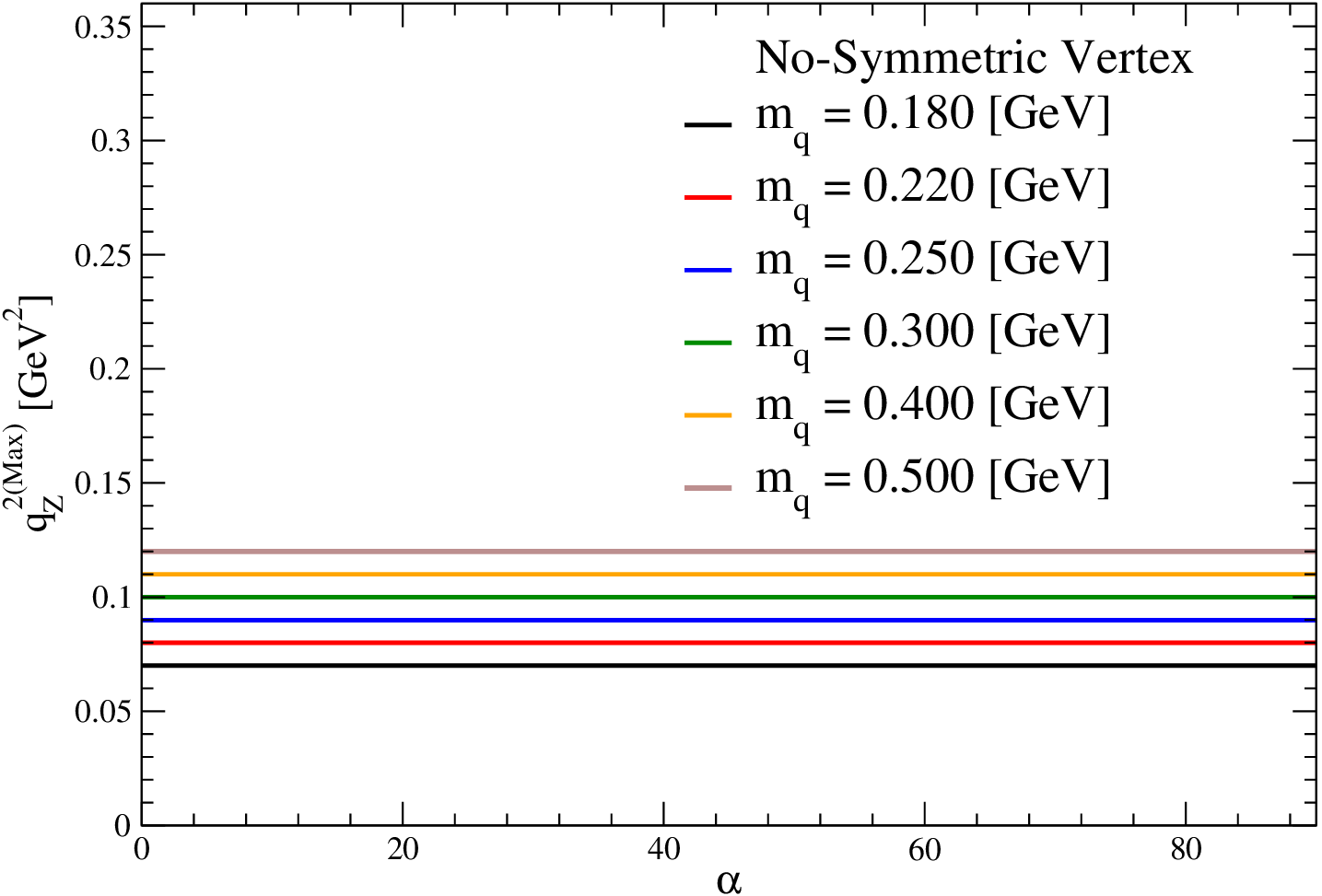}
	\caption{Left panel: The maximum momentum transfer $q^2$ for the 
		pion electromagnetic form factor, in function of the frame angle, for 
		for each quark mass.
		Right panel:~
		The same of the panel left, but for the non-symmetric vertex model \cite{deMelo2002}.
		For both figures, the maximum $q^2$ is constant for for each quark mass, 
		and frame independent.
		} 
	\label{fig10}  
\end{figure}

  In Figure~\ref{fig10}, we present the maximum momentum transferred to the non-valence
   terms as a function of the variation of the angle $\alpha$,
    with quark masses in the range $\in [180, 500]\text{ MeV}$.
     Consequently, for each quark mass, we obtain a fixed momentum, 
     as seen in the figures above, and this fixed momentum is independent of
      the reference frame, depending instead on the constituent quark mass 
      scale adopted in the present work.
      
      This figure summarizes what can be observed in the previous figures, 
      namely, Figs. 7 through 9, showing the contributions of the zero modes to the elastic pion form factor as a function of quark masses for both models used in the present work~\cite{deMelo1999,deMelo2002}.
       As observed in \Cref{fig7,fig8,fig9} in the present work, the zero-mode contribution $F_\pi^{(Z)}(Q^2)$ exhibits a pronounced peak at a specific momentum transfer scale, denoted as $Q^2 = q_{Z,\text{max}}^2$.
      
       A key feature highlighted by these results is that 
       the maximum of the zero modes contribution always occurs at the exact same $Q^2$ value, 
       regardless of the choice of the reference frame angle $\alpha$ (for $\alpha \neq 0^\circ$).
 
 \section{summary}
 \label{summary}
 
 The frame dependence parameterized by the angle $\alpha$ (which governs $q^+$ 
 components) controls the \emph{magnitude} and overall weight of the non-valence contributions.
  When $\alpha = 0^\circ$ (corresponding to $q^+ = 0$, or the Drell-Yan Breit frame condition),
   the zero-mode contribution vanishes entirely~\cite{deMelo1999,deMelo2002,Leao2023,Moita2021} 
   for pseudoscalar meson.
  
   As $\alpha$ increases toward $90^\circ$, the magnitude of $F_\pi^{(Z)}(Q^2)$ grows significantly. However, the functional dependence on the momentum transfer $Q^2$, 
  dictated by the internal loop
   integration over the constituent quark propagators and vertex functions, remains governed by
    the intrinsic dynamics of the bound state rather than the observer's frame.

    The position of the peak $q_{Z,\text{max}}^2$ is strictly determined by the internal mass scales of the Light-Front Constituent Quark Model (LFCQM), namely the constituent quark mass $m_q$ and the regulator mass $m_R$ (appearing in the Bethe-Salpeter vertex functions). 
    As demonstrated in the Figures above, varying $m_q$ shifts the value of $q_{Z,\text{max}}^2$, 
    but for any fixed quark mass and vertex model (symmetric and non-symmetric), 
    the peak location remains completely invariant across all frame angles $\alpha$.

 The zero modes represent virtual pair creation and annihilation processes ($\bar{q}q$ vacuum fluctuations relative to the light-front surface) embedded in the Feynman triangle diagrams.
 
  The probability and momentum space density of these non-valence configurations peak at a characteristic internal momentum scale dictated by the binding dynamics of the pion. When probed by an external virtual photon carrying momentum $Q^2$, the overlap integral maximizes at this characteristic scale, producing a stable, frame-independent peak position $q_{Z,\text{max}}^2$.

  In conclusion, while the magnitude of the zero-mode contributions 
  strongly depends on the reference frame ($\alpha$) due to the breaking 
  and subsequent restoration of rotational symmetry, the localization of the 
  maximum effect at a specific $Q^2$ is an intrinsic dynamical property.
  
  It reflects the characteristic momentum scale of non-valence quantum 
  fluctuations inside the pion bound state as modeled by the LFCQM.
    
  Finally, consideration is being given to extending the behavior of zero modes, or the pair terms observed here for pseudoscalar mesons (specifically the pion in the present case), 
  to other pseudoscalar mesons, as well as to vector mesons.  
  We highlight that this extension, a continuation of the present work to the spin-1 case (where zero modes are even more complex due to polarization), 
  already shows preliminary indications consistent with the invariance of $q_{Z,\max}^2$ found for the pion case.

\begin{center}
	{\bf ACKNOWLEDGMENTS}
\end{center}
 
This work was supported in part by 
the Conselho Nacional de Desenvolvimento Cient\'{i}fico e Tecnol\'{o}gico~(CNPq), Brazil, 
(Grant No. 351403/2025-6).  
Funda\c{c}\~{a}o de Amparo \`{a} Pesquisa do Estado de S\~{a}o Paulo (FAPESP), Brazil, 
(Grant No.2023/09539-1), and FAPESP thematic project (Grant 
No. 2024/17816-8); 
 and was also part of the projects, 
Instituto Nacional de Ci\^{e}ncia e Tecnologia-Nuclear Physics and Applications~(INCT-FNA,MCTI),
Process No. 408419/2024-5, and CNPq, CAPES and FAPERJ, Brazil.     
\\ \\


.

\vspace{.80cm}

\vspace{1.0008750cm}

\vspace{1.0008750cm}

\clearpage 
\newpage 

\onecolumngrid

 \bibliography{zeromodesv1}

\end{document}